\RequirePackage{fix-cm}

\documentclass[smallextended]{svjour3}       

\usepackage{amsfonts, indentfirst, latexsym, bm,graphicx,amsmath,amssymb,mathrsfs,verbatim,hyperref,pdfpages}
\usepackage{caption}
\usepackage{subcaption}
\usepackage{color}
\usepackage[T1]{fontenc}
\usepackage{dsfont}
\usepackage{latexsym}
\usepackage{url}
\usepackage{array}
\usepackage{tikz}
\usepackage{tkz-berge}
\usepackage{tkz-graph}
\usepackage{pgfplots}
\usepackage{graphicx}
\usepackage{layout}
\usepackage[titletoc,title]{appendix}
\usepackage{bbm} 
\usepackage{algorithm,algorithmic}
\numberwithin{equation}{section}

\newcommand{\Pl}{\mathbb P}
\newcommand{\E}{ \mathbb E}

\def\E{\mathbb{E}}
\def\Var{\mathrm{Var}}
\def\Cov{\mathrm{Cov}}
\usepackage{chngcntr}
\counterwithout{equation}{section} 
\usetikzlibrary{arrows, automata,positioning,calc,shapes,decorations.pathreplacing}
\def\VR{\kern-\arraycolsep\strut\vrule &\kern-\arraycolsep}
\def\vr{\kern-\arraycolsep & \kern-\arraycolsep}
  \colorlet{greencolor}{green!50!black}
  \colorlet{textcolor}{red}
  \colorlet{tancolor}{orange!80!black}
  \colorlet{bluecolor}{blue}
  \definecolor{plotcolor1}{rgb}{0,0.447,0.741}
\definecolor{plotcolor2}{rgb}{0.741,0,0.447}
\definecolor{plotcolor3}{rgb}{0,0.741,0.294}
\definecolor{plotcolor4}{rgb}{0.741,0.294,0}
\definecolor{plotcoloraux}{rgb}{0.447,0.447,0.447}
\tikzset{plotstyle1/.style={color=plotcolor1,solid,line width=1.0pt}}
\tikzset{plotstyle2/.style={color=plotcolor2,densely dashed,line width=1.0pt}}
\tikzset{plotstyle3/.style={color=plotcolor3,dotted,line width=1.0pt}}
\tikzset{plotstyle4/.style={color=plotcolor4,loosely dashed,line width=1.0pt}}
\tikzset{auxlines/.style={color=plotcoloraux,solid,line width=0.5pt}}
\newcommand{\markersize}{0.7pt}
\tikzset{discretemarkers/.style={mark=*,mark options={solid},mark size=\markersize}}
\usepackage{makecell} 

 \journalname{
Queueing Systems: Theory and Applications}
\begin{document}

\title{A Survey of Parameter and State Estimation in Queues
}

\titlerunning{A Survey of Parameter and State Estimation in Queues}        

\author{ Azam Asanjarani        \and Yoni Nazarathy  \and Peter Taylor
}


\institute{ Azam Asanjarani\at The University of Auckland
              \\
              \email{azam.asanjarani@auckland.ac.nz}           
                          \and
          Yoni Nazarathy \at The University of Queensland
          \and
          Peter Taylor \at The University of Melbourne       
}

\date{Received: date / Accepted: date}

\maketitle

\begin{abstract}
We present a broad literature survey of  parameter and state estimation for queueing systems. Our approach is based on various inference activities, queueing models, observations schemes, and statistical methods. We categorize these into branches of research that we call estimation paradigms. These include: the classical sampling approach, inverse problems,  inference for non-interacting systems, inference with discrete sampling, inference with queueing fundamentals, queue inference engine problems, Bayesian approaches,  online prediction, implicit models, and  control, design, and uncertainty quantification. For each of these estimation paradigms, we outline the principles and ideas, while surveying key references. We also present various simple numerical experiments. In addition to some key references mentioned here, a periodically-updated comprehensive list of references dealing with parameter and state estimation of queues will be kept in an accompanying annotated bibliography.

\end{abstract}
\keywords{Queueing inference; queueing parameter estimation; inverse problems; queue inference engine; queueing prediction.}

\newpage
\section{Introduction}
\label{intro}

Queues occur in retail, health-care, telecommunications, manufacturing, road traffic, social justice systems, call centres and many other environments. To aid in understanding such systems, mathematical queueing models have been studied and employed for over a century. Such models allow researchers and practitioners to predict congestion and delay behaviour based on assumptions about the underlying stochastic processes. The field has grown together with the field of applied probability and constitutes a significant part of the world of stochastic operations research. Indeed, queueing phenomena are both fascinating and important to understand from a practical perspective. 

The basic building block of most queueing analysis research involves a {\em queueing model}. As an example, consider a model called the M/D/1 queue. In such a model, customers arrive to a server according to a Poisson process. The server processes customers one at a time, taking a fixed deterministic time, $m$, for each customer and idles when there are none left.  When a customer arrives to see the server busy, the customer queues. 
In this model, there are only two parameters of interest: $\lambda$ the arrival rate (customers per time unit), and $m$ the service time. 
If $\lambda m > 1$ then the average number of customers that arrive during a service is greater than one and the queue will build up over time without bound. However, if $\lambda m < 1$ then the system will settle down to a stochastic equilibrium. 

Each customer experiences a waiting time (which may be $0$ if arriving to an empty queue) and a sojourn time which is the customer's total time in the system (waiting time + service time). In the $\lambda m < 1$ regime, it makes sense to analyse equilibrium mean waiting times and other performance measures. There are multiple service disciplines that the server may use such as first come first served (FCFS), random order of service, or other disciplines. The service discipline generally affects the distribution of the waiting time, but does not affect the mean waiting time as long as the server doesn't idle when customers are present. Queueing theory thrives on results for models such as M/D/1. For example, by setting $\rho =\lambda m<1$, the mean waiting time of an arbitrary customer for this model is
\begin{equation}
\label{eq:md1mean}
m \frac{\rho}{2(1-\rho)}.
\end{equation}

Formulas such as this  immediately lead to elementary system insights. First observe that as $\rho \to 1$ the mean waiting time grows without bound. For example with $\rho = 2/3$  the mean time a customer waits in the system is equal to the time it takes the customer to be served. However for higher values of $\rho$, the mean waiting time of a customer is longer than the service time.

Results such as this  are the cornerstone of queueing theory and analysis. However, how can queueing theory be employed? An immediate answer is to use queueing analysis for arriving at general insights about real world systems. For example, an insight gained from the result presented above is that mean waiting times are of the order $\frac{\rho}{1-\rho}$, as $\rho \to 1$. 


Such insights have helped with the design of computing systems, telecommunication systems, manufacturing systems, health-care operations and more.
With an abundance of queueing  models, one may wish to go further than just providing insight. Indeed, there is the  opportunity to use these models to predict, manage and design explicitly. This requires using queueing models that are realistic in the context of actual systems and the associated data collection processes. As an example, say that we know that a telecommunications switch takes exactly $1$ millisecond to handle a packet ($m = 0.001$ seconds). Now under the Poisson assumption for packet arrivals, for $\lambda \in (0,1000)$ we know that  the mean waiting time can be computed  using \eqref{eq:md1mean}. For example if $\lambda = 700$ then with an M/D/1 model, the mean waiting time is about $1.166$ milliseconds. Such conclusions require data collection for estimation of $m$, $\lambda$, and verification of the suitability of the model.

Such an application of queueing results generates a variety of questions associated with statistical analysis. In an actual system, how would we measure $\lambda$? Or what about the Poisson arrival assumption, is it sensible and supported by data? Further, what if different aspects of the queueing system were observed, not necessarily giving us a full indication of all the underlying processes. How would we then fit a queueing model to the system?

It turns out that, while there are thousands of papers dealing with queueing theory and analysis, there are far fewer papers dealing with these types of estimation problems. In fact, this state of affairs was identified as early as 1965 by David R. Cox in \cite{cox1965some} where he stated,

\vspace{5pt}

\setlength{\leftskip}{0.5cm}
\setlength{\rightskip}{0.5cm}

{\em  
There are a very large number of papers on particular probabilistic models for queues and, by comparison, extremely few papers on the corresponding problems of statistical analysis. When a simple mathematical model is investigated primarily to get qualitative insight into the behavior of queueing systems, the statistical problems are not very relevant. When, however, there is the possibility of quantitative application, or when a practical congestion problem is tackled by rather empirical methods,
non-trivial statistical problems arise.
}

\setlength{\leftskip}{0pt}
\setlength{\rightskip}{0pt}

\vspace{5pt}

Our purpose in this survey is to present  results that are available and  work that has already been done.  Such problems were considered quite early with the seminal work of Clarke \cite{clarke1957maximum}, the survey \cite{bhat1997statistical} by Bhat, Miller, and Rao, and an updated survey with the same title \cite{bhat1987statistical} by Bhat and Rao.  Since these were published, there has been a significant body of additional work. We survey both the classical queueing estimation work and more recent results in the current paper.

\subsection*{Structure of this survey}

This survey is structured as follows: We start in Section~\ref{sec:frameBackground} by describing the general framework. This includes outlining in Section~\ref{sec:Problems Solved} a variety of problems which we call {\bf inference activities}. We then present a brief illustration of queueing models in Section~\ref{sec:models}. This section is geared towards those that haven't been exposed to queueing theory. We then go on to present what we define as {\bf observation schemes} in Section~\ref{sec:obs}. In Section~\ref{sec:methods} we illustrate some of the complexities involved with parameter estimation in queues. The survey continues in Section~\ref{sec:settingAndActivities} where we lay out  the various {\bf estimation paradigms} and outline some of the key contributions in the literature. See Table~\ref{Tab1} for an overview. We conclude in Section~\ref{conclusion} where we outline a few broad areas that have received less attention in the literature. The computational examples that we present in Section~\ref{sec:methods} are also accessible via the GitHub repository \cite{Nazar_github}. A (periodically updated) annotated bibliography aiming to contain an exhaustive list of references in the area is in \cite{asanjarani2017parameter}.

\section{Framework and Background}
\label{sec:frameBackground}

Research that deals with inference in a queueing model usually has a number of characteristics that describe the type of inference and the type of modelling that is involved. We can classify a paper in this area according to four general attributes:

\begin{enumerate}
\item [(i)]{\bf The inference activity} that is performed, for instance, parameter estimation, state estimation, hypothesis testing, or sample size planning. 
\item [(ii)]{\bf The models} employed such as an M/M/1 model, an open queueing network model, or an M/G/$\infty$ model. 
\item [(iii)]{\bf The observation scheme} used, such as whether a continuous record of data is available or just data observed at certain time points.  
\item [(iv)]{\bf The statistical methods and principles} used, for instance,  likelihood based methods, moment matching, Bayesian inference, or non-parametric inference. 
\end{enumerate}

We discuss (i) -- (iv) in the subsections below. 

\subsection{Inference Activities}
\label{sec:Problems Solved}

Performing inference on queues can have different objectives. Here are some common activities and their objectives:

\begin{enumerate}
\item {\bf To find the parameters of a model.} In this case, we believe or assume that a real queueing system for which we have data behaves according to a specific model. The task is to estimate parameters of such a model. It could for example be to estimate the arrival rate $\lambda$ for the M/D/1 queue discussed in the introduction.
The majority of the work that we survey in this paper deals with this type of activity. 
\item {\bf To select a suitable model based on data.} The act of choosing a model for a scenario is often performed without reference to data. 
However, in certain cases, we may want to incorporate data into the model selection process. For example we may want to test if interarrival times to a queueing system are independent and exponentially distributed to decide if a Poisson arrival process assumption is suitable. 
There hasn't been much work on this type of activity (in the context of queues) to date. We survey the few papers that we found. 
\item {\bf To plan observation schemes and experiments.}  In classical statistical contexts, elementary considerations in design of experiments involve determining the number of samples to take, the various treatment classes, and stratifying and randomising subjects. In a queueing context, there is an additional complication due to the fact that a queueing system is a dynamic process evolving over time. Most studies are necessarily observational studies. Many such schemes involve indirect methods for observing the quantities of interest. Important considerations involve efficiency of information retrieval and understanding such things as sampling bias. Only a few of the papers that we survey, deal with such an activity and we believe that there is room for further research on this area.
\item {\bf To carry out state prediction or filtering}. In (1) above we discussed estimating parameters of models. However, in many practical situations, a parameterised model is already present and the question is about the state. Given past partial observations, it is of interest to either estimate the full state in the past or predict future states. Such problems may often be tackled via black box methods such as neural networks and/or hidden Markov models. However in our context, the methods are based on actual queueing models. 

%
\item {\bf Adaptive control.}  The process of estimating states and parameters and the process of controlling the system are often decoupled. However, in certain cases, one may make decisions on-line while parameter and state estimation is ongoing. This is the case of adaptive control. In general, methods from reinforcement learning where the parameters are unknown, and partially observable Markov decision processes where the state is unknown present a variety of techniques for dealing with such problems.  However,  a few selected papers deal with such problems utilising the queueing structure. A related concept is robust control which is not exhaustively covered in this survey.
\end{enumerate}


\subsection{Queueing Processes and Models}
\label{sec:models}

We now briefly illustrate key aspects of queueing theory.  Our purpose is to present the reader with a taste of key phenomena, models, and quantities involved. Hence, this section is not about inference but is rather about the mathematical (stochastic) models of queues. Specialists in queueing theory may wish to skip this section as it contains elementary material.

As illustrated in the introduction, a model like M/D/1 may be used to predict expected waiting times, mean queue lengths and other measures in a system subject to Poisson arrivals. We now generalise this model to a broader class called the GI/G/1 queue. Other special cases are the M/G/1 and the M/M/1 models. All of these models are single-server queueing models (hence the ``1'' in the name). The difference between them lies in the probabilistic assumptions imposed on the arrival process. In all these models the interarrival times are assumed i.i.d.\ (independent and identically distributed). In the ``GI'' case they follow an arbitrary distribution while in the ``M'' case they follow an exponential distribution and in the ``D'' case, take a deterministic value. Similar comments apply to the service time process. More general queueing models allow dependence between interarrival times of the arrival process, and rarely between service times, which is less natural from a modelling point of view.

\paragraph{Exogenous Processes:} As model inputs, consider the following two basic {\em exogenous processes}: the arrival process and the service time process. 
The {\em arrival process} may be described by $A(t)$ where $t$ is a continuous time variable. Here $A(t)$ counts the number of arrivals during the time interval $[0,t]$. This process is exogenous because in the basic suite of models it is not considered to be affected by the internal dynamics of the queueing system. An alternative representation is via a sequence $T_1,T_2,\ldots$ that marks the arrival times of customers.

Like the arrival process, the {\em service time process} is usually assumed not to be influenced by the internal dynamics of the queue, hence it is an exogenous process. It is naturally described by the sequence, $S_1,S_2,\ldots$ where $S_n$ is the service time of the $n$th customer arriving to the queue. However, we can also define, 
\begin{equation}
\label{eq:sOft}
S(t) = \sup\{n ~:~ \sum_{i=1}^n S_i \le t \},
\end{equation}
as the ``service time analog'' of $A(t)$. We need to keep in mind that as time progresses, there are periods where the system is empty. Hence $S(t)$ is not necessarily the number of customers served during $[0,t]$. We discuss this in more detail below. 

\paragraph{Endogenous Processes:} Given a realisation of $\{A(t),~t \ge 0\}$ and $\{S_n\}_{n=1}^\infty$ together with some initial conditions, the essence of queueing modelling is the description and analysis of  {\em endogenous processes} that evolve. These include: 
\begin{description}

\item The {\em system size process}, $Q(t)$. This process specifies the number of items in the system at time $t$. Note that it is often called the {\em queue length process} even though it includes the customers being served (if any), as well as any customers  waiting in the queue.
\item The {\em waiting time sequence}, $\{W_n\}_{n=1}^\infty$. Here $W_n$ is the waiting time before entering service for the $n$th customer.
\item The {\em workload process}, $V(t)$. This process determines the volume of work in the system at any time $t$. It is sometimes called the {\em virtual waiting time process} as it indicates how long a customer arriving at time $t$ will need to wait (under the FCFS regime).
\item The {\em departure process}, $D(t)$. This process counts the number of departures (service completions) from the system during $[0,t]$.
\item The {\em busy period sequence}, $\{B_n\}_{n=1}^\infty$. A {\em busy period} is a duration of time during which the server is busy. It starts at time $\tau$ when $Q(\tau^-)=0$ and $Q(\tau) = 1$, with $A(\tau) - A(\tau^-) = 1$ due to a customer arrival. It then ends in the first time $\tilde{\tau} > \tau$ such that $Q(\tilde{\tau}) = 0$.
\end{description}

There are various ways to define the functional relationship mapping the exogenous processes and initial conditions to the above endogenous processes. One such simple example is based on the customer conservation equation,
\[
Q(t) = Q(0) + A(t) - D(t).
\]
This equation is useful if we describe the endogenous departure process, $D(t)$ in a different manner. To do so, we observe that $S(t)$  in \eqref{eq:sOft} determines how many customers could potentially be served during $[0,t]$ if there was always a customer present. Now also define the {\em idle-time process},
\[
I(t) = \int_0^t {\mathbf 1} \{ Q(u)=0 \} du,
\]
where ${\mathbf 1}\{\cdot\}$ is the indicator function. As a consequence, we can represent the departure process via the composition, 
\begin{equation}
\label{eq:dOft}
D(t) = S\big(t -I(t) \big) = S\Big( \int_0^t {\mathbf 1} \{ Q(u)>0 \} du \Big).
\end{equation}
Given initial conditions there is a unique endogenous process $Q(t)$ satisfying,
\begin{equation}
\label{eq:basicQ}
Q(t) = Q(0) + A(t) - S\Big( \int_0^t {\mathbf 1} \{ Q(u)>0 \} du \Big),
\end{equation}
see \cite{chen2013fundamentals} for details. A solution of $Q(\cdot)$ for \eqref{eq:basicQ} also describes $D(\cdot)$ via \eqref{eq:dOft}. In our context, equations \eqref{eq:dOft} and \eqref{eq:basicQ} serve the purpose of illustrating that the exogenous processes, $A(\cdot)$ and $S(\cdot)$, can be used to construct the endogenous processes $D(\cdot)$ and $Q(\cdot)$. Similarly the workload process $V(\cdot)$ can be expressed in terms of the exogenous processes via
\begin{equation}
\label{eq:workloadEq}
V(t) = V(0) + \sum_{i=1}^{A(t)}S_i - \int_0^t {\mathbf 1} \{ V(u)>0 \} du.
\end{equation}
Note that the integrals in \eqref{eq:basicQ} and \eqref{eq:workloadEq} are identical because $Q(t) > 0$ if and only if $V(t) > 0$.

Assume $Q(0) = 0$ and define the sequences $u_n$ for $n=1,2,\ldots$ and $v_n$ for $n=0,1,2,\ldots$ with $v_0 = 0$, such that,
\begin{equation}
\label{eq:lindleyR}
u_n = \inf \{t>v_{n-1} ~:~ Q(t)>0 \},
\qquad
v_n = \inf\{ t> u_n ~:~Q(t) = 0 \}.
\end{equation}
Then $B_n = v_n-u_n$ is the duration of the $n$th busy period. An analogous definition can be constructed when $Q(0) > 0$. This shows how the endogenous process $\{B_n\}$ can be constructed based on the exogenous processes.

In a similar spirit, the classic Lindley recursion,
\[
W_{n+1} = \max \{ W_n +  S_n - (T_{n+1}-T_n ), 0 \},
\]
determines $W_n$ based on the primitive sequences, $\{T_n\}$ and $\{S_n\}$. Here in agreement with $Q(0)=0$ we initialize the recursion with $W_1 = 0$. See for example \cite{boxma2007queues} for a modern treatment.

\paragraph{Stability and Traffic Intensity:} The notion of stability and a parameter known as traffic intensity appears in almost all queueing models. A canonical example is a GI/G/1 queueing system with arrival rate $\lambda$ and mean service time $m$. Such a system may behave differently depending on the value of $\rho = \lambda m$. If $\rho < 1$ then queues are stochastically stable, meaning that (under regularity conditions on the interarrival and service time distributions) as $t \to \infty$ the queue length process $Q(t)$ and waiting time processes $\{W_n\}$ converge to limiting distributions. Similarly if $\rho > 1$ then there is not enough capacity in the system to serve the incoming traffic and as $t$ grows, queues grow without bound almost surely. Further, at the critical value $\rho=1$, there is not a limiting distribution, however the queue ``grows at a slower rate'' as attested by the fact that the system empties infinitely often (yet with heavy tailed gaps between such instances). See for example \cite{foss2004overview}. In all these cases, it is clear that the \textit{offered load}, $\rho$, also known as the \textit{traffic intensity}, is a key quantity.

\paragraph{Probabilistic Performance Measures:} Queueing theory assumes a stochastic description of the exogenous processes and endeavours to determine stochastic descriptions of the endogenous processes. The types of stochastic processes involved include Markov processes, diffusion processes, L\'evy processes, and the analysis often involves associated limiting results for these types of processes. Frequently it is not possible to obtain a full description of the probability law of the endogenous processes, and we settle for summary measures such as the stationary queue length distribution or the stationary mean waiting time. 

An example of a key result in classical queueing theory is the Pollaczek-Khinchin (P-K) formula for M/G/1 queues. Here the arrival rate is $\lambda$, the service distribution has Laplace-Stieltjes transform $G(s)$ with mean $m$ such that $\rho = \lambda m < 1$. One version of the P-K formula gives an expression for the probability generating function $K(z)$ of the steady state queue length, whose random variable we denote by $Q$. In this case P-K, reads,
\[
K(z) = 
\sum_{k=0}^\infty z^k \, {\mathbb P}(Q = k) = 
(1-\rho) \frac{(1-z) G\big(\lambda(1-z)\big)}{G\big(\lambda(1-z)\big) - z},
\qquad
\mbox{for}
\qquad
|z| \le 1.
\]
In the M/D/1 case $G(s) = e^{-sm}$ and hence,

\begin{equation}
\label{eq:md1PGF}
K(z) = (1-\rho ) \frac{ (1-z)  e^{-\rho (1-z)}}{e^{-\rho  (1-z)}-z}.
\end{equation}

Now the stationary mean queue length can be computed by taking the first derivative and evaluating the limit as $z \to 1$. Further, the second factorial moment can be computed by taking the second derivative and evaluating the limit as $z \to 1$. From these we get the M/D/1 mean and variance,

\begin{equation}
\label{eq:MD1eQ}
\mathbb{E}[Q]=\frac{2 - \rho}{2}  \frac{\rho}{1-\rho}
,
\quad
 \operatorname{Var}(Q)= \frac{12-\rho(18+\rho(\rho-10))}{12} \frac{\rho}{(1-\rho)^2}.
\end{equation}

We use these formulas in Section~\ref{sec:methods}, illustrating statistical methods. When the service time distribution is exponential, the system is called an M/M/1 queue, in which case $G(s) = (1+ms)^{-1}$ and thus,
\[
K(z) = \frac{1-\rho}{1-\rho z},
\]
which is the generating function of a geometric distribution with support $0,1,2,\ldots$, mean $\mathbb{E}[Q] = \rho/(1-\rho)$, and variance $\operatorname{Var}(Q) = \rho/(1-\rho)^2$. Comparing with \eqref{eq:MD1eQ}  we see that as $\rho \to 1$, the mean queue length is reduced by a factor of almost~$2$ and the variance by a factor of almost $4$. 
 
\paragraph{Little's Law:} Under general conditions, we can show that queueing systems in steady state satisfy Little's Law:
\begin{equation}
\label{eq:little}
\ell =\lambda \tau,
\end{equation}
where $\ell$ is the steady state number of customers in the system, $\lambda$ is the arrival rate of
customers through the system, and $\tau$ is the mean steady state sojourn time of each customer. The interpretation of ``system'' can change depending on context. For example, Little's Law holds for the waiting room of customers, or the total service facility (the waiting room together with customers in service).

To illustrate the use of Little's law we  can use the expectation in \eqref{eq:MD1eQ} to obtain \eqref{eq:md1mean}. For this observe that the mean sojourn time $\tau$ is the sum of the mean service time $m$ and the mean waiting time, $w$. Now solving $\frac{2 - \rho}{2} \frac{\rho}{1-\rho}
 = \lambda (w + m)$ for $w$, we obtain \eqref{eq:md1mean}.


\paragraph{Stochastic Process Limits in Queueing Theory:} We often wish to consider models with more complexity than the M/G/1 queue. This can include either a more detailed description of how customers, servers, and allocation policies interact, or more general assumptions on the endogenous processes. In such cases, exact results such as the P-K formula are often not attainable. Nevertheless, much of the effort in queueing theory research over the past few decades has focused on more involved models. A key approach is to use stochastic process limits which give a theoretical basis for approximating the endogenous processes via limiting processes which are easier to handle.

The basic building blocks of such mechanisms involve {\em fluid limits} and {\em diffusion limits}. The idea of a fluid limit is to consider only the first order deterministic approximation of associated processes. For example, the arrival counting process $A(t)$, can be approximated by the function $\bar{A}(t) = \lambda t$, and similarly for $S(t)$. Such a view of queueing systems ignores the randomness but often captures the essence of the system, especially when considering stability or the behavior at large. For example, a fluid limit approximation of a GI/G/1 queue starting with $Q(0) = q_0$ customers is,
\[
\bar{Q}(t) = \max\{q_0 - (\frac{1}{m} - \lambda) t, 0\}.
\]

Such an approximation indicates that if $\rho \ge 1$ then the queue will not ``drain'' whereas if $\rho < 1$ then at approximately $t= m \,q_0 /(1-\rho)$ the queue will hit zero. This is a good approximation if the process starts with a large initial state $q_0$. Fluid limit based approximations, clearly ignore the subtle stochastic variations that play a key role in results such as the P-K formula presented above. 

A second order refinement considers the fact that deviations between exogenous processes such as $A(t)$ and their fluid limit $\bar{A}(t)$ can often be approximated via diffusion processes. Here the idea is to consider a sequence of systems indexed by $n=1,2,\ldots$ and construct processes such as,
\[
\breve{A}_n(t) = \frac{A(nt) - \bar{A}(nt)}{\sqrt{n}}.
\]

It then turns out that under mild assumptions on $A(\cdot)$, the sequence of processes $\{\breve{A}_n(\cdot)\}$ converges weakly to a drift-less Brownian motion process. Carrying out such approximations on all or some of the exogenous processes, then allows us to derive approximations for the endogenous processes.

A very fruitful framework occurs when we also scale the parameters of the exogenous processes such that $\rho_n \to 1$ from below. This suite of limiting regimes yields {\em heavy traffic} approximations for the endogenous processes and their performance measures. Other forms of scaling such as the Halfin-Whitt regime also known as the quality and efficiency driven (QED) regime, are also very popular. See for example \cite{whitt2002stochastic} for an overview.

To get a feel for the strength of such methods consider approximating the waiting time distribution of a GI/G/1 queue as $\rho_n \to 1$, see Corollary~7.5, \cite{asmussen2008applied}. Referring to our M/D/1 example, while we can use the P-K formula to compute clean expressions such as \eqref{eq:md1mean}, computing the actual distribution of the stationary waiting time is not as simple. Nevertheless, a heavy traffic approximation such as that in Corollary~7.5 in \cite{asmussen2008applied} implies that with $\rho \approx 1$, the waiting time distribution is approximately exponential with parameters that depend on the mean and variance of the interarrival time $T$ and the service time $S$. Specifically for the M/D/1 model
%
\[
\lim_{\rho \to 1} \Pl(W_\rho >x) =  \exp\{ - \rho^2 \, \frac{2( 1-\rho)}{m \rho } x \} = \exp\{ - \rho^2 \frac{1}{\E[W_\rho]} x \},
\qquad
\text{for}
\quad x \ge 0.
\]
This is an exponential distribution with mean $\rho^{-2}\E[W_\rho] \approx \E[W_\rho]$.
For more general GI/G/1 queueing systems we are not able to compute $\E[W_\rho]$ explicitly, nevertheless a heavy traffic limit approximation such as Corollary~7.5 in \cite{asmussen2008applied} is very powerful because all that is needed is the mean and variance of the interarrival and service times.

\paragraph{Additional Branches of Queueing Theory:} We should also mention several other sub-fields of queueing theory that have allowed us to obtain results for the endogenous processes. One major branch is {\em matrix analytic methods} which involves modelling the queueing processes with structured Markov chains which are amenable to algorithmic computation of certain performance measures, see \cite{latouche2012matrix}. Another branch deals with networks of queues, where even though the system is often quite high dimensional and complex, under general assumptions, one may often show that the stationary distribution possesses a {\em product form} structure. See for example the classic book \cite{kelly2011reversibility}, or more modern treatments in \cite{bramson2008stability}. A third branch involves {\em tail asymptotics} where results dealing with probabilities such as $\Pl(W > x)$ are approximated for large $x$.

\subsection{Observation Schemes}
\label{sec:obs}

Having explored elements of queueing theory and related processes, we now present a possible classification of observation schemes for inference. When characterising methods and results associated with queueing inference, a first step is to consider which processes are observed and which are not. For example, we may observe the queue length process, the workload process, the arrival process, the service process, or some combination thereof. A second step is to consider how well these processes are observed, for example fully or only at discrete time points. Table~\ref{tab:obs} contains a few major descriptors that we shall use to organize the discussion below.

\begin{table}[h!]

     \begin{tabular}{ |>{\centering\arraybackslash}  m{1.5cm} |>{\arraybackslash}  m{9.5cm} 
 | }
    \hline
\textbf {Notation} & \textbf{ \hspace{2.5 cm}   Observation Scheme  }
       \\
    \hline
(\bf{F}) &{\bf Full observation}: This is the situation where the relevant random processes are observed continuously, possibly during some bounded time interval.
 \\   
 \hline           

{\bf (DI)} & {\bf Discrete intervals}: This is the situation where one or more of the random processes
is sampled at some discrete time points. An example can be the case where in a single server queue we sample every $\Delta$ time units. Hence the data is $\{Q(i \Delta) \}$, for  $i = 1, \cdots, n$.
 \\   
 \hline           
{\bf (IO)} & {\bf Input and output process observation}: This is the situation where only the arrival process and departure process are sampled. An example is an infinite server queue where we see customer arrival times and departure times but do not know with certainty how to match arrivals and departures.
   \\   
 \hline           
 {\bf (P)} & {\bf Probing}: This is the situation where customer journeys of only selected (often manually injected) customers are observed. An example can be a communication network where there is a major traffic flow and we are injecting occasional probe customers to measure behaviour.
   \\   
 \hline           
 {\bf (T)} & {\bf Transactional observations}: This is the situation where only service commencements and completions are observed together with an indication of server businesses. For example, such an observation scheme may occur in an automatic teller machine where queues are unobservable but server activity is being logged. 
   \\   
 \hline           
 {\bf (IP)} & {\bf Independent primitives}: In this scheme, we don't actually observe the queueing process, but rather observe some of the random variables that construct it, often with a pre-specified number of observations.

   \\   
 \hline           
        
 \end{tabular}
 \caption{Different kinds of observation schemes.}
 \label{tab:obs}
\end{table}

\begin{figure}[h]
\centering
\includegraphics[scale=0.5]{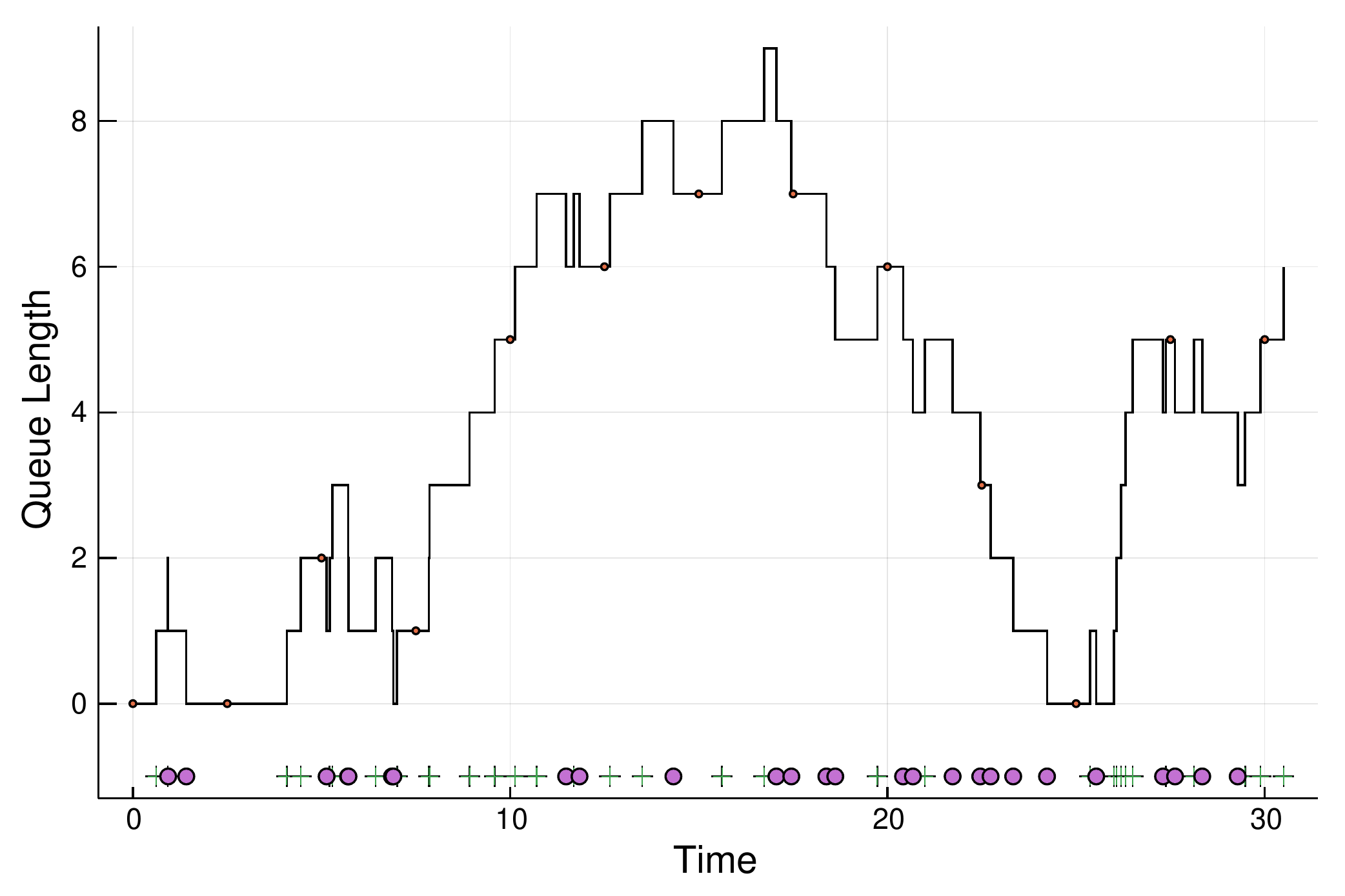}
\caption{\label{fig:queueTrajectory}
{\small A simulated sample path of $Q(t)$ for an M/M/5 system over $t \in [0,30]$. If observed using (F) then the full trajectory of $Q(t)$ is available. If observed using  (DI) then discrete samples are collected every $\Delta = 2.5$ time units. If observed using (IO) then the sequence of arrivals to the queue (marked with $+$ symbols) and the sequence of service completions (marked with $\bullet$ symbols) is available as data. The figure does not present samples using  (P), (T) or  (IP)  cases. 
}}
\end{figure}
   
Note that it is possible that a particular inference activity could potentially be classified under more than one descriptor. Our purpose is not to fully categorize a situation via the classification but rather indicate the general nature of the information available.
In certain cases and when using certain models, without considering  initial conditions on the queue length, observation schemes (F) and (IO) are equivalent. In fact, based on the queue length process, we can determine the arrival and departure times uniquely for single-server models where the customer order is deterministic. However, finding the queue length from the arrival and departure epochs requires information about the initial value of the queue length.

As an illustrative example let us consider an M/M/$5$ queue. This is a system with $5$ servers, Poisson arrivals, and exponential service time for each customer. That is, if $Q(t) \le 5$ then all customers in the system are served simultaneously during time $t$, and further when $Q(t) > 5$ then $Q(t)-5$ customers are waiting for service. Figure~\ref{fig:queueTrajectory} illustrates a simulated trajectory of such a system with the purpose of highlighting several types of data sequences, relating to the different observation schemes above.

\subsection{Statistical Methods}
\label{sec:methods}

The vast field of statistics provides methods for carrying out a variety of tasks. 
In this survey we focus mainly, but not solely, on estimation in which case either parameters of models, state estimates, or non-parametric estimates are produced based on collected data. Such estimation can be carried out either in the classic frequentist setting or a Bayesian setting. 
The reader should keep in mind that many methods of elementary statistics are typically presented in the context of independent and identically distributed (i.i.d.) random variables. Adapting such methods to queueing inference, often requires considering the dependencies and dynamics of the underlying queueing models. We now present an example.

We return to the M/D/1 queue and explore statistical inference under the (DI) observation scheme. Here, the queue length process is sampled $n$ times; every $\Delta$ time units. Taking the first observation at time $\Delta$, the sample can then be represented as,
\begin{equation}
\label{eq:sampleQ1}
X = \Big(Q(\Delta),Q(2\Delta),Q(3\Delta),...,Q(n \Delta)\Big).
\end{equation}

We explore two alternative inference activities, both assuming the underlying system is in steady-state. First assume that we simply wish to estimate the steady state mean queue length (an endogenous performance measure) and obtain confidence intervals for our estimate. Later we consider parameter estimation for the arrival rate $\lambda$.

\paragraph{Confidence Intervals for $\E[Q]$:} Using classic statistical formulas a naive approach would be to estimate the mean queue length via the sample mean,
\[
\overline{X} = \frac{1}{n} \sum_{i=1}^n Q(i\,\Delta),
\]
and then to continue to obtain a $95\%$ confidence interval
\begin{equation}
\label{eq:classicStatsCI}
(\overline{X} - 1.96 \frac{S}{\sqrt{n}}, \overline{X} + 1.96 \frac{S}{\sqrt{n}}),
\end{equation}
for $\E[Q]$ with,
\begin{equation}
\label{eq:sFormula}
S = \sqrt{\frac{1}{n-1} \sum_{i=1}^n (Q(i \Delta) - \overline{X})^2}.
\end{equation}

If $X$ is an i.i.d.\ vector of observations that are normally distributed with mean $\theta$, then $\sqrt{n} (\overline{X} - \theta)/S$ follows a $t$-distribution with $n-1$ degrees of freedom which, for large or even reasonably sized $n$, is approximately a standard normal distribution. Then $1.96$ is approximately the $0.975$th quantile of a standard normal distribution and this yields the confidence interval formula \eqref{eq:classicStatsCI}. Even if the observations are not normally distributed, the central limit theorem ensures that $\overline{X}$ has an approximate normal distribution if $n$ is large.

While we may get away with assuming stationarity, queuing processes generally exhibit strong dependence over time. There are versions of the central limit theorem that apply to dependent sequences (see for example \cite{asmussen2008applied}, page 30). However, there is still the problem of estimating the variance of the limiting normal distribution. In particular, if $\Delta$ is not large, the covariances strongly influence the variance of $\overline{X}$ via,
\[
\Var(\overline{X}) = \frac{1}{n} \Var(Q) +  \frac{1}{n^2}  \sum_{i \neq j} \Cov(Q(i\Delta),Q(j \Delta)),
\]
where $Q$ represents a generic random variable of the queue size in steady state and $\Var(Q)$ is as in \eqref{eq:MD1eQ}.
Also, similar calculations yield,
\[
\E[S^2] = \Var(Q) +  \frac{1}{(n-1)n}  \sum_{i \neq j} \Cov(Q(i\Delta),Q(j \Delta)).
\]
Hence due to the covariance terms, the estimation of the standard deviation via \eqref{eq:sFormula} may be heavily biased. This jeopardizes the validity of the confidence interval \eqref{eq:classicStatsCI}.  We demonstrate this effect via a numerical experiment.  

Assume a ground truth with mean service time $m=1$ and $\lambda = 0.9$, and hence $\rho = 0.9$. This implies the steady-state unknown mean queue length is $4.95$ as per \eqref{eq:MD1eQ}. To estimate it, we could take $n=100$ samples and consider different scenarios when $\Delta$ is in the range $10,20,\ldots,300$. For each value of $\Delta$, we simulate $M=10^4$ Monte-Carlo simulations of the queue, letting it ``warm up'' for $10^3$ time units each time (this sets it close to ``steady state''). Each simulation run samples the queue as per \eqref{eq:sampleQ1}. We then estimate the coverage probability of the resulting confidence interval via,
\[
{\cal C}_\Delta = \frac{1}{M} \sum_{i=1}^M {\mathbf 1} \Big\{4.95 \in \big(\overline{X}-1.96 \frac{S}{\sqrt{n}},
\overline{X} + 1.96 \frac{S}{\sqrt{n}}\big) \Big\}.
\]

\begin{figure}[h]
\centering
\includegraphics[scale=0.5]{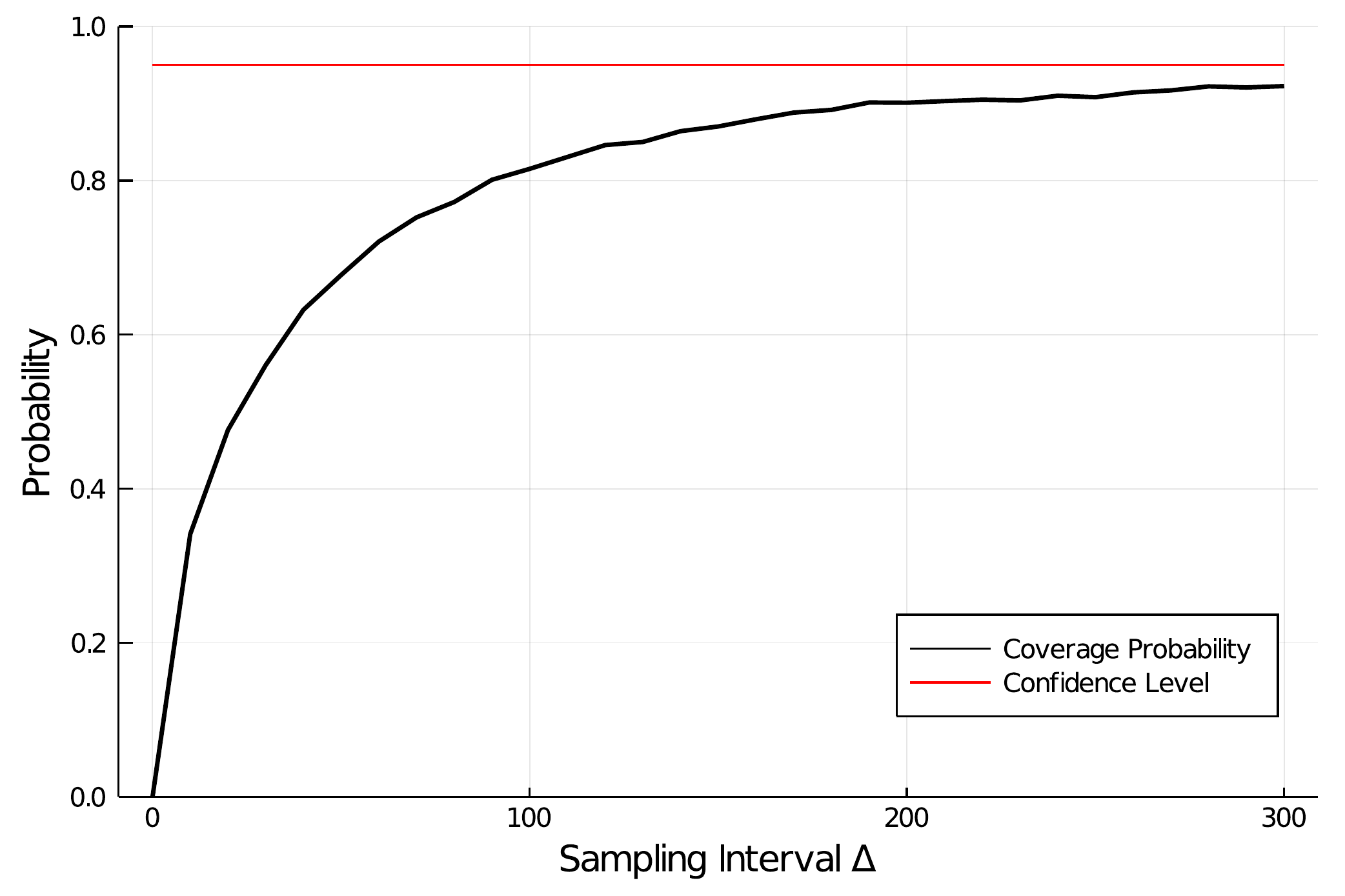}
\caption{\label{fig:ciCoverage}
{\small Coverage probability obtained by using the confidence interval \eqref{eq:classicStatsCI} with increasing $\Delta$ and $n=100$.
}}
\end{figure}

The estimates are plotted in Figure~\ref{fig:ciCoverage}. As expected, as $\Delta$ grows the coverage probability agrees with the i.i.d.\ case. However, for small $\Delta$ we see a big discrepancy between the actual coverage probability and the desired $95\%$. Hence for small $\Delta$ we see that the classic confidence interval formula \eqref{eq:classicStatsCI} breaks down. 

Clearly, in our construction we used a naive confidence interval that assumes no covariance between $Q(i \Delta)$ and $Q(j \Delta)$ for $i \neq j$ and this is the cause of the error. We should mention that there has been extensive work on such estimation for time-series where the samples are not i.i.d., see for example \cite{brockwell1991time}. Still, in the context of queueing, one may often try to use the explicit model structure, as opposed to assuming arbitrary covariance structures as is common in the time-series literature. We survey examples of this in the sequel.

 \paragraph{Estimating $\lambda$:} Say now that under the same observation scheme, we know that $m=1$ and we wish to estimate $\lambda$. For this we can develop an estimator based on $\E[Q]$ from \eqref{eq:MD1eQ}. If we set $\overline{X}$ on the left hand side of $\E[Q]$ in \eqref{eq:MD1eQ}, and  solve for $\lambda$, we obtain $\lambda = 1+\overline{X} \pm \sqrt{1+\overline{X}^2}$. We can take the negative sign ensuring that $\rho \in (0,1)$. This isn't hard to check for any positive $\overline{X}$. Hence our estimator is,
 \begin{equation}
\label{eq:lambdaMD1est}
\hat{\lambda} = 1+\overline{X} - \sqrt{1+\overline{X}^2}.
 \end{equation}

Let us now evaluate the quality of this estimator using the mean squared error criterion,
\[
\text{MSE} = \E[(\hat{\lambda} - \lambda)^2],
\]
and determine how $\Delta$ affects the MSE. We can also reason about the limiting MSE as $\Delta$ becomes large and $n \to \infty$. For large $\Delta$ it is reasonable to assume that $Q(i \Delta)$ and $Q(j \Delta)$ for $i \neq j$ are independent observations of the stationary queue length random variable $Q$. Then from the central limit theorem, $\overline{X}$ is approximately normally distributed with mean $\E[Q]$ and variance $\Var(Q) /n$ and the limiting mean square error is,
\begin{equation}
\label{eq:limitMSE}
\widetilde{\text{MSE}} =  \int_{-\infty}^\infty (1+z - \sqrt{1+z^2} - \lambda)^2 \frac{1}{\sqrt{\Var(Q)/n}} \phi\Big( \frac{z - \E[Q]}{\sqrt{\Var(Q)/n}} \Big) d z,
\end{equation}
where $\phi(\cdot)$ is the standard normal density. Note that we are not able to evaluate the right hand side of \eqref{eq:limitMSE} analytically, but rather use numerical integration.

In our case with the ground truth of $\lambda = 0.9$ and $m=1$, using \eqref{eq:MD1eQ} we have $\E[Q] = 4.95$ and $\Var\big(Q \big) = 23.7825$. For $n=100$, \eqref{eq:limitMSE} yields $\widetilde{\text{MSE}} = 0.010074^2$.  Also the central limit theorem typically ``kicks in'' for moderate values of $n$. Hence for $n=100$, assuming independence, normality effectively holds. However, for smaller $\Delta$, the situation is different as we present in this numerical experiment.

As before, we consider different scenarios for $\Delta$  in the range $10,20,\ldots,300$. For each value of $\Delta$, we simulate $M=10^5$ Monte Carlo simulations of the queue, letting it ``warm up'' for $10^3$ time units. Over each simulation run we estimate $\hat{\lambda}$ and then for each value of $\Delta$ we estimate the root MSE via,
\begin{equation}
\label{eq:mseEst}
\text{RMSE}_\Delta =\sqrt{ \frac{1}{M} \sum_{i=1}^M (\hat{\lambda} - 0.9)^2}.
\end{equation}

\begin{figure}[]
\centering
\includegraphics[scale=0.5]{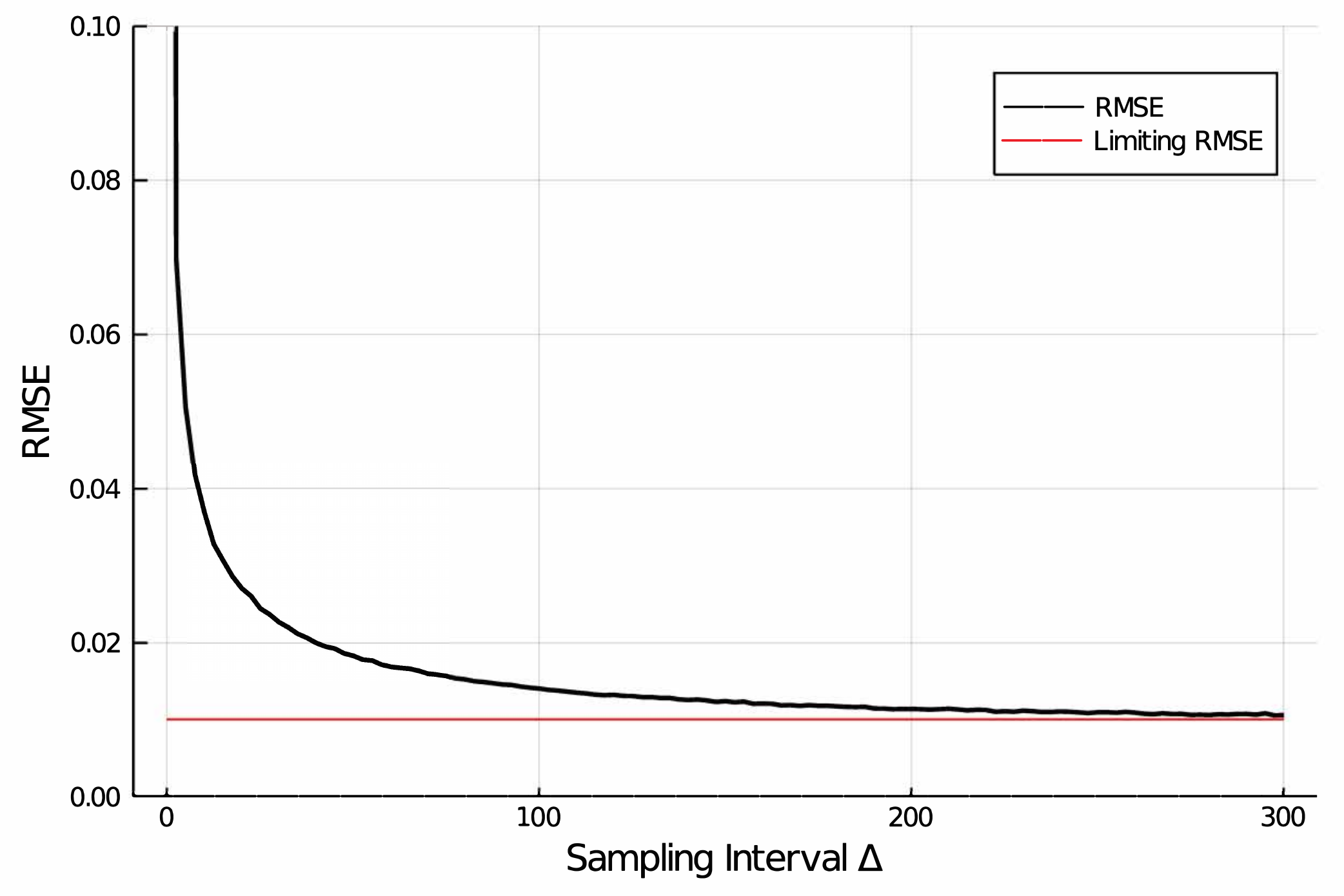}
{\small \caption{\label{fig:mse}
{\small The root MSE estimate \eqref{eq:mseEst} with increasing $\Delta$, $n=100$.}}
}
\end{figure}

The resulting Monte-Carlo RMSE estimates are plotted in Figure~\ref{fig:mse} and we indeed see that as $\Delta$ grows, our limiting approximation, $\sqrt{\widetilde{\text{MSE}}}$ holds.
\paragraph{The Broader View of Statistics:} The two examples above illustrate that classical statistical methods can break down when carrying out inference for queues if dependencies are ignored.
Nevertheless, when designing queue inference procedures it is important to be aware of the vast set of tools developed in classic and contemporary statistics. 

For example, the two most common ways of finding the estimated parameters are the method of moments and maximum likelihood estimation (MLE). The most important benefit of the method of moments is that it is usually fast and often non-iterative. However, like MLE, the method of moments often yields non-unique estimators and both methods are difficult to apply when the number of parameters is large. Further, in many cases, method of moment estimators are less efficient than MLE. 

Likelihood based approaches view the observed data under a certain model. MLE provides estimates for model parameters which yield the largest likelihood of the observed data. The widespread use of maximum likelihood is due to the asymptotic properties of the MLE. For instance,  model estimates coming from  the MLE are optimal in terms of asymptotic variance. Another advantage of using MLE is that many model selection procedures  such as AIC (Akaike's information criterion) and BIC (the Bayesian information criterion) are  based on MLE. The computational effort required for the MLE is its major drawback. Note that even with i.i.d. data, MLEs are often biased estimators, however under general conditions, MLEs are asymptotically unbiased. Nevertheless, with 
dependent data, such as waiting times, one often ends up with 
biased parameter estimates, see \cite{basawa1996maximum} for more details.

In queueing inference we are often faced with incomplete data. One generic useful tool for this scenario is the \textit{EM algorithm}. It is a way of maximizing the likelihood that is quite effective for estimating parameters of models with some hidden or incomplete data. The name ``EM algorithm'' stems from the alternating application of an \textit{expectation} step (E-step) and a \textit{maximization} step (M-step) that yields a successively higher likelihood of the estimated parameters. See for example \cite{asmussen1996fitting}, in the context of \textit{phase-type} (PH) distributions, often closely related to queueing models. The EM algorithm is also broadly applicable for \textit{hidden Markov models} (HMMs) which can be used to represent certain queueing inference scenarios. 

An alternative is the Bayesian approach where we consider parameters to be random variables. We assign a prior probability distribution  to each unknown parameter. Applying Bayes' rule we update the parameter belief distribution from the prior to the posterior. 

Another important avenue is non-parametric analysis. For stochastic models such as queues ideas were developed by Gr\"{u}bel \cite{grubel1989stochastic}. This method is based on considering the queueing model as a mapping from the exogenous processes to the endogenous processes. Then, under certain conditions such as continuity of the mapping, applying differentials can provide information about changes to the endogenous processes based on changes of the exogenous processes. Further local properties of the functional can give  valuable information about the robustness, consistency, and asymptotic  properties of the estimators, see \cite{grubel1992functional}.   

\section{Various Estimation Paradigms}
\label{sec:settingAndActivities}

We now move to the heart of the survey. We present and summarize a variety of results and methods. In considering queueing inference problems, there are several dimensions at play. These include:
\begin{itemize}
\item The physical/real-world problem being investigated and the goals associated with the application.
\item The queueing model (or class of models) being used. See Section~\ref{sec:models}.
\item The observation scheme. See Section~\ref{sec:obs}.
\item The type of statistical methodology being used. See Section~\ref{sec:methods}.
\end{itemize}
The interplay of these aspects are  woven into a specific domain that we call an {\bf estimation paradigm}. 

 Having studied the broad literature, we decided to partition the scope of this field into ten paradigms. Each paradigm shares a specific sub-field of research. One major characteristic of each paradigm is the typical observation scheme. We summarize the paradigms and their typical observation schemes in Table~\ref{Tab1} and survey them in the subsequent subsections. Note that for each estimation paradigm we refer only to a few selected key references. The reader can find a more complete list of references dealing with queueing inference in the annotated bibliography \cite{asanjarani2017parameter}. 


\begin{table} 
\begin{tabular}{ |>{\arraybackslash}  m{5.6cm}  |>{\centering\arraybackslash}  m{1.5cm} |>{\arraybackslash}  m{3.45cm} | }
    \hline

\centering{  \textbf {Estimation paradigm} }&
\centering{ \textbf{Typical obs scheme}}&
 \textbf{$\qquad\,$Key references}
\\  

 \hline           
{\footnotesize
{\bf The Classical Sampling Approach}:
This is the situation where either endogenous or exogenous processes are sampled, or both. 
The actual number of samples might be fixed or might be a random variable determined endogenously.
}
&
{\bf (F)}, {\bf (IP)}
 & \cite{bhat1997statistical},
\cite{bhat1987statistical},
\cite{basawa1988large},
\cite{clarke1957maximum},
\cite{thiruvaiyaru1991estimation},
\cite{rodrigo1999large},
\cite{basawa2008parameter},
\cite{wang2015maximum},
\cite{schruben1982some},
\cite{zheng2000some},
\cite{ke2009comparison},
\cite{kiessler2009technical}

 \\ 
 \hline
 
    {\footnotesize
{\bf Inverse Problem Estimation}:
This is a situation where certain attributes of the system are observed and these observations are used to infer model parameters.
} 
&
{\bf (P)}
&\cite{chen1994parameter}, 
\cite{sharma1998estimating}, 
\cite{alouf2001inferring}, 
\cite{hei2005light},
 \cite{hei2006model}, 
\cite{nam2009estimation}, 
\cite{comert2009queue},
 \cite{novak2009determining},  
\cite{baccelli2009inverse},  
\cite{sharma1998estimating}, 
\cite{pin2010statistical},
\cite{heckmuller2009reconstructing},  
 \cite{kauffmann2012inverse}, 
\cite{antunes2014probing}, 
\cite{antunes2016estimation}, 
 \cite{kim2018data},
\cite{bingham1999nonparametric},
\cite{hansen2006nonparametric},
\cite{pitts1994nonparametric}

   \\   
 \hline           
   
{\footnotesize
{\bf Inference for Non-Interacting Systems:} This paradigm deals with models where customers don't interact such as the M/G/$\infty$ queue and generalisations.
}
 &
{\bf (IO)}, {\bf (DI)}
&
   \cite{reynolds1975covariance}, \cite{brown1970m},  \cite{blanghaps2013sojourn}, \cite{grubel2011matchmaking}, \cite{goldenschluger2016nonparametric}, \cite{goldenshluger2018m}, \cite{goldenshluger2019nonparametric}, 
   \cite{edelmann2014nonparametric}, \cite{schweer2015nonparametric},
   \cite{schweer2020nonparametric},
    \cite{hall2004nonparametric},  \cite{bingham1999non}, 
   \cite{pickands1997estimation}
   
   \\   
 \hline           
   
{\footnotesize   
{\bf Inference with Discrete Sampling}: This paradigm focuses on cases where systems are sampled discretely over time.
}
&
{\bf (DI)}
 & \cite{bladt2005statistical},
\cite{ross2007estimation}, \cite{mcvinishconstructing},
\cite{ravner2019estimating}, \cite{mandjes2019hypothesis}, 
\cite{duffie2004estimation}, 
\cite{den2017convergence}
\\   
 \hline           

{\footnotesize   
{\bf Inference with Queueing Fundamentals}: This paradigm describes situations where queueing theory fundamentals aid parameter and state estimation. The most prominent example is the use of Little's law.  }
&
{\bf (F)}
& \cite{glynn1989indirect}, \cite{glynn28estimating}, \cite{duffy2009estimating},
\cite{carson1980conservation}, \cite{glynn1986central}, \cite{nozari1988estimating}, \cite{glynn1993estimating},
\cite{kim2012estimating},
\cite{mandjes2017detecting}

\\
 \hline           
{\footnotesize
{\bf Queue Inference Engine Problems}:  This paradigm deals with a branch of problems where transactional observations are recorded and the trajectory of the queue within a given cycle is inferred.  
}
&
{\bf (T)}
& \cite{larson1990queue}, \cite{bertsimas1992deducing}, \cite{daley1992exploiting}, \cite{daley1993two}, \cite{jones1999inferring}, \cite{jones1994efficient},  \cite{daley1997estimating}, \cite{daley1998moment}, \cite{dimitrijevic1996inferring}, \cite{mandelbaum1998estimating}, \cite{fearnhead2004filtering}, \cite{frey2010queue}, \cite{jones2012remarks},   \cite{park2011analysis},   \cite{heckmuller2011reconstructing}

\\   
 \hline           
{\footnotesize   
{\bf Bayesian Approaches}: In most of the Bayesian work to date the parameter estimation utilises known queueing performance analysis formulas, considering their posterior distributions given a sensible choice of priors.
} &
{\bf (IP)}
&
\cite{bolstad2016introduction},
\cite{armero1985bayesian}
\cite{armero1994bayesianInf}, 
\cite{armero1994bayesian},
 \cite{armero1994prior},
\cite{insua1998bayesian}, 
 \cite{ausin2004bayesian},
\cite{ramirez2010bayesian},
\cite{mcgrath1987subjective}, 
\cite{mcgrath1987subjectiveII},
\cite{sutton2011bayesian},
\cite{neal2003slice},
\cite{sutton2011bayesian},
\cite{wang2016maximum},
\cite{thiruvaiyaru1992empirical},
\cite{conti1999large},

 \\   
\hline 
\footnotesize{
{\bf Online Prediction}: In this paradigm, we observe the states up to a  given time and make prediction about future states. 
}
&
{\bf (F), (DI)}
&
\cite{cohen1982single}
\cite{stanford1983optimal},
\cite{woodside1984optimal},
\cite{pagurek1988optimal},
\cite{whitt1999predicting},
\cite{ibrahim2011wait},
\cite{thiongane2016new}

\\
\hline
{\footnotesize   
{\bf Implicit Models}: This paradigm deals with new developments combining data-science and queueing theory where queue-like models are introduced without explicitly modeling every component of the system.
}& 
{\bf (F)}
&

\cite{senderovich2014queue},
\cite{van2004workflow},
\cite{senderovich2015discovering},
\cite{senderovich2015queue},
\cite{dong2015stochastic},
\cite{au2009predicting},
\cite{zhang2002workload},
\cite{whitt2012fitting},
\cite{dong2015using}

\\   
 \hline           
{\footnotesize   
{\bf Control, Design, and Uncertainty Quantification}: This paradigm deals with work related to parameter and state estimation where control and design decisions based on inferred values are to be made. 
}& 
&
\cite{hernandez1983adaptive},
\cite{krishnasamy2018learning},
\cite{asanjarani2019role},
\cite{armony2005impact},
\cite{dinh2014architecture},
\cite{bandi2015robust}
\\
 \hline           
\end{tabular}
 \caption{Key references for different estimation paradigms and their most relevant observation schemes.}
  \label{Tab1}
\end{table}


\newpage

\subsection{The Classical Sampling Approach} 
\label{likelihood}

When considering parameter estimation for queueing systems, a natural first step is to consider observation scheme (F), see Table~\ref{tab:obs}, where all the data is available. In that case, one may think that there aren't any challenges because we can simply employ the state of the art parameter estimation methods for the queueing primitives (interarrival and service times). This isn't far from the truth when we have large samples available, however, for small samples there are some technical complications. These complications have driven most of the classic research on parameter estimation of queues. Much of this work is summarized in the last queueing estimation survey by Bhat et. al, \cite{bhat1997statistical}. See also the earlier survey, \cite{bhat1987statistical} by Bhat and Rao.

As an example, consider a single server queue where we observe the sequence of interarrival times $\{A_1,\ldots,A_{n_a}\}$ and the sequence of service times $\{S_1,\ldots,S_{n_s}\}$, where these sequences are i.i.d., independent of each other and the sample sizes $n_a$ and $n_s$ are fixed. We can then naturally estimate $\lambda$ and $\mu$ via,
\begin{equation}
\label{eq:lamMuNaive}
\hat{\lambda}=\frac{n_a}{\sum_{i=1}^{n_a} A_i},
\qquad\text{and} \qquad
\hat{\mu}=\frac{n_s}{\sum_{i=1}^{n_s} S_i}.
\end{equation}

However when observing a queue, $n_a$ and $n_s$ are often not fixed and can be dependent on the sequences $\{A_i\}$ and $\{S_i\}$. Complications may arise, not only due to censoring, but also due to the dependency structure of the various quantities. 
Here are some possibilities:
\begin{itemize}
\item We may sample the system for a fixed duration $[0, T]$ in which case both the number of arrivals and the number of service completions are dependent random variables.
\item We may sample a fixed number of arrivals, $n_a$, in which case the observation time, $T$, and the number of service completions are dependent random variables. 
\item We may sample a fixed number of service completions, $n_s$, in which case the observation time, $T$, and the number of arrivals are dependent random variables.
\item We may sample using some other similar scheme (such as a fixed number of transitions) which will again imply that other quantities are random variables. 
\end{itemize}
In each of these cases the estimation procedure exhibits what we refer to as {\em endogenously determined sample sizes}. That is, the total number of either interarrival, service times, or both is a random quantity resulting from the model.  This aspect drove much of the early research on parameter estimation of queues and is well described in \cite{basawa1988large}. In fact, the subtle problems that arise in such cases were considered in one of the first parameter estimation papers for queues by Clarke in 1957, \cite{clarke1957maximum}. 

The work in \cite{clarke1957maximum} focused on parameter estimation for a stationary M/M/1 system where the parameters are the arrival rate $\lambda$ and the service rate $\mu$. When sampling an M/M/1 queue for a fixed duration $[0,T]$,  it is difficult to obtain a simple likelihood expression for the unknown parameters $\lambda$ and $\mu$. Hence a more creative sampling scheme was proposed where a set duration, $\tilde{T}$ is determined and sampling takes place for as long as the busy time of the server is less than $\tilde{T}$. In such a case, standard properties of the M/M/1 queue imply that the likelihood can be written as,
$$
L(\lambda,\mu ~;~ \text{data}) = \big( 1-\frac{\lambda}{\mu} \big)\, e^{-\mu \tilde{T}-\lambda T_{n_s}}\,\mu^{n_s-\nu} \,\lambda^{n_a+\nu}K (n_a,n_s,\nu, T_{n_s}),
$$
where $K$ does not depend on the unknown parameters $\lambda$ and $\mu$. This expression is useful because the (full observation) data is summarized via the statistics $n_a$, $n_s$, $\nu$, and $T_{n_s}$.  As defined previously, the statistics $n_a$ and $n_s$ are the number of arrivals and number of service completions (only this time endogenously determined via the sample). The statistic $\nu$ is the initial queue size. Finally, the statistic $T_{n_s}$ is the time of the last service completion during $[0,\tilde{T}]$. This structure of the likelihood allows one to maximise with respect to $\lambda$ and $\mu$ given measurements of the sufficient statistics. Further, there is also the (minor) extra added benefit that the initial queue length, $\nu=Q(0)$ can yield more information for this observation scheme. 

As exemplified by the results of \cite{clarke1957maximum}, the (F) observation scheme, while simple, still entails some interesting challenges. However, when considering larger sample sizes, the subtle issues associated with the construction of MLEs and similar estimators are not as crucial. Nevertheless, a significant body of literature has handled such queuing inference problems. For instance, in \cite{thiruvaiyaru1991estimation} the problem of estimation for tandem queues was discussed as a special case of Jackson networks. Further, in \cite{rodrigo1999large}, the case of a general G/G/1 retrial queue was considered. Here, the flow of repeated attempts can be non-Markovian and the system is observed until there is a fixed number of departures. Then, in \cite{basawa2008parameter} estimation of the parameters of GI/G/1 queues was studied where only the incomplete information of the differences between service and interarrival times was observed. Also, in \cite{wang2015maximum}, MLEs for service demands in closed queueing networks with load-independent
and load-dependent stations were proposed.

Going back to the simple estimator \eqref{eq:lamMuNaive}, even in the situation where $n_a$ and $n_s$ are fixed, there may be anomalies in the inference process. For example, combining $\hat{\lambda}$ and $\hat{\mu}$ from \eqref{eq:lamMuNaive} we have an estimator for the offered load,
\begin{equation}
\label{Eq:naive_rho}
\hat{\rho}=\frac{\hat{\lambda}}{\hat{\mu}},
\end{equation}
which may appear straightforward. However, in \cite{schruben1982some}, Schruben  and Kulkarni showed that for an M/M/1 queue if we wish to use $\hat{\rho}$ to compute the steady state mean queue length, some unexpected behaviour may occur. The ratio of the estimated traffic intensity to the
true traffic intensity has an F distribution with $2n_s$ and $2n_a$ degrees of freedom. Further, they showed that this estimator has undesirable sampling properties. For example, even when we restrict the estimated workload to be strictly less than one (for instance, by re-sampling for the case that $\hat{\rho}\geq 1$), the expected value of the plug-in estimator $\frac{\hat{\rho}}{1-\hat{\rho}}$ 
for the average number of customers is infinite. That is,
\begin{equation}
\label{eq:rhoIsBad}
\E[\frac{\hat{\rho}}{1-\hat{\rho}}\,\, \mathbf{1}{\{\hat{\rho}<1\}}]= \infty.
\end{equation}

These types of results indicating {\em anomalies in inference} are useful to keep in mind when a practitioner estimates primitives as inputs into basic queueing models such as M/M/1, but also for more complex discrete event simulation models. That is, in simulation modelling practice one often considers system primitives as inputs into a complex discrete event simulation. Then a discrete event (say agent-based) simulation model can be used for performance analysis. The take-home message from a simple result such as \eqref{eq:rhoIsBad} is that such problems can also occur in much more complex models.

The results from \cite{schruben1982some} were generalized in \cite{zheng2000some} where alternative estimators for the limiting expected number of customers in the queue (and several other performance measures) were constructed. 
These estimators require  the analyst to choose a value $\rho_{0} < 1$.  Under the assumption that $\rho < \rho_{0}$, the estimator has finite mean and finite mean square error.
Further, in \cite{ke2009comparison}, similar estimators were considered including the consideration of bootstrap based confidence intervals as well as other statistical aspects. A third notable paper, dealing with this aspect of queueing estimation is Kiessler and Lund \cite{kiessler2009technical}, where the authors proposed and analyzed two alternative estimators for $\rho$ in M/G/1 queues.

The first estimator discussed in \cite{kiessler2009technical}  uses the sample average of the work arriving into the system during $[0, T]$, 
 \begin{equation}
\label{eq:rhoWork}
 \hat{\rho}_{\text{work}}=\frac{\sum_{i=1}^{N(T)}S_i}{T}= \hat{\rho}\, \frac{\sum_{i=1}^{N(T)}A_i}{T},
 \end{equation}
where $N(T)$ is the total number of customers arriving up to time $T$, $\{A_i\}$ and $\{S_i\}$ are as above, and $\hat{\rho}$ is similar to the estimator in \eqref{Eq:naive_rho} with $n_a$ and $n_s$ equalling $N(T)$.

 The second estimator uses the proportion of time during which the server is busy: 
 \begin{equation}
\label{eq:rhoVirtual}
 \hat{\rho}_{\text{virtual}}=\frac{\int_{0}^{T} \mathbf{1}{\{V(u)>0\}}
 \,du}{T},
\end{equation}
 where $V(t)$ is the workload process.
The paper showed that \eqref{eq:rhoVirtual} is an asymptotically unbiased estimator and further provided analysis of asymptotic means, biases, and variances.

\subsection{Inverse Problem Estimation}
\label{prob}

In general an inverse problem arises in a situation where we observe endogenous processes and need to estimate or predict parameters of exogenous processes. As a simplest example consider a stationary M/D/1 queue and the mean waiting time given in \eqref{eq:md1mean}. Assume we know the value of the mean service time $m$ and observe the waiting times of $n$ customers. If the observed sample average waiting time is $\overline{W}$, then we can estimate $\rho$ via the equation,
\begin{equation}
\label{eq:md1Est2}
\overline{W} = m \frac{\rho}{2(1-\rho)}.
\end{equation}
Solving for $\rho$ we obtain an estimator,
\begin{equation}
\label{eq:md1Est3}
 \hat{\rho}  = \frac{\overline{W}}{\overline{W}+m/2}.
\end{equation}
 
Much of the work dealing with inverse problems has focused on the (P) observation scheme, see Table~\ref{tab:obs}. In certain situations, endogenous processes, such as $\{W_n\}$ yielding $\overline{W}$, are not directly observable, and we need to obtain observations by actively probing the system via artificial customers (packets). The prober usually chooses the sizes of probes and time epochs in which to send them. In this case we say that the probing is \textit{active}. However, sometimes  the probe sizes are determined by the selected application and associated network protocol such as transport control protocol (TCP). In this case we have {\em passive} probing. 

\begin{figure}[h]
\centering
\includegraphics[scale=1.5]{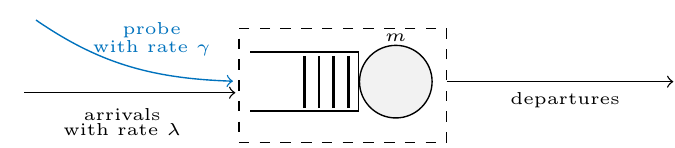}
\caption{\label{fig:probe}
{\small Probing a single server queue. Regular arrivals  arrive at Poisson rate $\lambda$. The probe stream is injected with rate $\gamma$. The service duration for both probes and regular packets has a mean of $m$.}}
\end{figure}

Probing is depicted in Figure~\ref{fig:probe}. In such a case, the probes slightly affect the system, and via their measurements we aim to solve an inverse problem.

Continuing with the M/D/1 example, assume we wish to estimate $\rho$. To do this the prober sends $n$ probes into the queue at the time points of a truncated Poisson stream with rate $\gamma$. Here  $\gamma$ should typically be quite small so as not to disturb the system. For each probing customer, the prober measures the sojourn times denoted via $\tau_1,\ldots,\tau_n$ with $\overline \tau_p$ denoting their average.
Then we can estimate $\overline W$ from the waiting times experienced by the probes via $\overline W = \overline \tau_p - m$. 
Then using equation \eqref{eq:md1Est2} again with $\rho = (\lambda + \gamma)m$,
we can solve for $\lambda m$ to obtain the estimator,
\begin{equation}
\label{eq:probEstimatorMD1}
\hat{\rho}_{\text{M/D/1 probes}} = \frac{\overline{\tau}_p - m}{
\overline{\tau}_p - m/2
} - \gamma \, m.
\end{equation}

As a comparison, consider  the case where the underlying system is a stationary M/M/1 queue with a known mean service time $m$, and arrival rate $\lambda$ which is unknown. In this case since the M/M/1 mean sojourn time
is $(m^{-1} - \lambda - \gamma)^{-1}$, a probing based estimator for the offered load is,
\begin{equation}
\label{eq:probEstimatorMM1}
\hat{\rho}_{\text{M/M/1 probes}} = \frac{\overline{\tau}_p - m}{\overline{\tau}_p} - \gamma\, m.
\end{equation}

Note that in practice, we don't always know if the underlying system is better modelled as an M/D/1 queue, an M/M/1 queue, or some other model. Still, for a given observations yielding $\overline{\tau}_p$, the estimators in \eqref{eq:probEstimatorMD1} and \eqref{eq:probEstimatorMM1} show that treating the system as an M/D/1 queue will yield a higher offered load estimate than treating it as an M/M/1 queue.

One of the first probing papers \cite{chen1994parameter} by Chen et. al. 
considered a FCFS M/D/1 system.  The  prober knows the arrival times, waiting times, and departure times of probes. The authors derived a tractable expression for the likelihood function of $\lambda$. This allows us to carry out maximum likelihood estimation.

In  the general case of G/G/1 queues,  assume that $\lambda$ and $\mu_1$ are the average arrival rate and service rate of local traffic, and $\gamma$ and $\mu_2$ are the average arrival rate and service rate of the probes. 
Then, the traffic intensity is 
\begin{equation}\label{eq:rho_M/G/1}
\displaystyle \rho=\frac{\lambda}{\mu_1}+\frac{\gamma}{\mu_2},
\end{equation}
which, in the case that $\rho <1$, is the stationary fraction of time during which the server is busy. This can form the basis of an estimator.
If $\gamma$, $\mu_1$ and $\mu_2$ from \eqref{eq:rho_M/G/1} are known then we can estimate $\lambda$ by using \eqref{eq:rho_M/G/1} together with an estimator like \eqref{eq:rhoVirtual}.

However, in reality the $Q(t)$ (or $V(t)$) is often not observed. So, in \cite{sharma1998estimating}
Sharma and Mazumdar introduced a method based on measuring the delay experienced with active probing.
They solved the  problem for the cases where the true arrival process is either Poisson or arbitrary and the probing process is Poisson, denoted by M$+$M/G/1 and M$+$G/G/1, respectively. Further, they extended the problem to cases where the service times are unknown as well as  queueing networks. This was the first paper that proposed an analytic approach to probing of queueing networks. 

In \cite{alouf2001inferring}, Alouf et. al.\ 
studied the case where  the system  has limited unknown capacity $c$. They considered Poisson arrivals and service times that are either exponential or deterministic and denoted the systems via M$+$M/M/1/$c$ and M$+$M/D/1/$c$. Their estimators for $c$ can be used to estimate system capacity.

When $\rho \ge 1$ a different approach can be used. In \cite{hei2005light} Hei et. al. observed that the ratio ${\cal R}$ between the mean interarrival and mean interdeparture time is given by
$$
{\cal R} =
\begin{cases}
1 \qquad \qquad \,\, \,\rho <1,\\
\frac{\lambda}{\mu_1}+\frac{\gamma}{\mu_2} \qquad \rho \geq 1.
\end{cases}
$$
In this case, an estimate of ${\cal R}$ based on averages of observed interarrival and interdeparture times can be used to estimate $\lambda$. 

When $\rho <1$, the ratio ${\cal R} =1$. This yields no information about the arrival rate and we need to consider the second order characteristics of the departure process. The authors suggested using an approximation of the squared coefficient of variation (SCV) of the inter-departure times to estimate the arrival rate of a D$+$M/G$_i$/1 queue where the notation $G_i$ indicates that the service distribution may differ between the customers and the probes. In follow up work \cite{hei2006model}, the authors extended the method to the case of M$+$M/G$_i$/1 queues.

Analysis for special cases of probing and development of new estimation methods was a very active area of research around the turn of the first decade of the current century. 
In \cite{nam2009estimation} Nam et. al.\ considered probing for parameter estimation of an M/G/1 queue where both the service rate $\mu=1/m$ and the input traffic load $\lambda$ are unknown. Their method estimates the available bandwidth (the residual processing capacity) based on what they call a minimal-backlogging method. 
Comert and Cetin in \cite{comert2009queue} considered the application of probing for real-time estimation of the number of vehicles (customers in a queue) in a signalised traffic intersection.  This is the case that only the position of the last probing customer in the queue is known. 
In \cite{novak2009determining}, the authors studied the convergence rate of an M/D/1 queue to its steady state as a function of the load. They attempted to use this performance measure for finding an adequate probe separation threshold.

In a significant paper \cite{baccelli2009inverse},  Baccelli, Kauffmann, and Veitch described how to apply probing methods for queueing networks. Following the initial work in \cite{sharma1998estimating}, the work in \cite{baccelli2009inverse} presented a comprehensive survey of probing methods to estimate parameters and design queueing networks. See also, \cite{pin2010statistical} where the authors dealt with network tomography, and further exploited the EM algorithms for multicast trees.

In \cite{heckmuller2009reconstructing} Heckm{\"u}ller and Wolfinger studied estimation of the arrival rate of a G/D/1 queue with probing where only the departure times are observed. Their method was constructed in a discrete time setting, approximating the numbers of customers arriving in time intervals by Gaussian random variables. They also investigated sequences of queues with possibly varying bottleneck capacity. In \cite{kauffmann2012inverse}, Kauffmann, suggested a new approach with zero probing overhead based on the theory of inverse problems for bandwidth sharing networks.
In \cite{antunes2014probing}, Antunes et. al.\ considered the problem of estimating the arrival rate and the service rate of  an M/G/1 queue with probing.  They also studied the time-varying M$_t$/G$_t$/1 queues in \cite{antunes2016estimation}. In \cite{kim2018data} Kim et. al.\ applied a data driven probing approach to provide a  high-fidelity simulation model for an arrival process to a clinic.

\paragraph{Non-parametric Problems:}
While the (P) observation scheme, see Table~\ref{tab:obs}, has attracted most of the attention in the realm of inverse problems, other types of settings are also relevant in practice. For example non-parametric estimation for the M/G/1 queue is considered in \cite{bingham1999nonparametric} based on busy periods and in \cite{hansen2006nonparametric} using the P-K formula. Based on workload observations, the authors of \cite{hansen2006nonparametric} constructed estimators both for the offered load $\rho$ and the (non-parametric) service time distribution. In earlier work, \cite{pitts1994nonparametric}, Pitts studied inference for GI/G/1 queueing models and laid foundations for non-parametric inference. Further non-parametric work appears in the context of infinite server queues, a topic which we cover next.

\subsection{Inference for Non-Interacting Systems }
\label{non-interact}

We now discuss inference associated with the M/G/$\infty$ queue and similar models. We call such systems {\em non-interacting} because customers do not affect each other in the queue. Almost any probabilistic analysis of an M/G/$\infty$ model is based on a transformation of the Poisson arrival process and this is why many M/G/$\infty$ results (and generalizations) are tractable. While there is not real ``queueing'' taking place in such a model, infinite server  systems naturally appear in applications as they describe a situation where incoming customers experience random delay. One example is a pedestrian crossing tunnel where pedestrians don't really interact, and the delay between the entry time and the exit time of each pedestrian are i.i.d. random variables.

Just as an illustration, we can compare the formulas for the auto-covariance function of the stationary queue-length for an M/M/1 queue and an M/M/$\infty$ queue, both with a mean service time of $1$.   From  \cite{reynolds1975covariance},  the auto-covariance is, 
\[
\text{Cov}\big(Q(0),Q(t)\big)
=
\begin{cases} \displaystyle
\frac{2\lambda (1-\lambda)}{\pi}\int_0^\pi \frac{(\sin \theta)^2 e^{-t(1+\lambda-2\sqrt{\lambda}\cos \theta)}}{(1+\lambda-2\sqrt{\lambda}\cos \theta)^3}d\theta
& \mbox{for M/M/1},\\
\lambda e^{- t} & \mbox{for M/M/}\infty.
\end{cases}
\]
As is evident having non-interacting customers (M/M/$\infty$) yields a much simpler formula. 

The analytic tractability of M/M/$\infty$ queues (and M/G/$\infty$ queues for that matter) goes beyond the auto-covariance function. Many performance measures have closed form expressions involving the arrival rate and the service time distribution $G(x)$. This has motivated several authors to consider inverse problems in various settings. Specifically observations schemes of types (DI), (IO), and (T) given in Table~\ref{tab:obs} have been studied.

\paragraph {The (IO) Observation Scheme:} A neat initial M/G/$\infty$ result from Brown in \cite{brown1970m} deals with the  transformation of the service time distribution
\begin{equation}
\label{eq:mgHG}
H(x) = 1-(1-G(x))e^{-\lambda x}.
\end{equation}
As is easy to see (Lemma 2 of \cite{brown1970m}), $H(\cdot)$ happens to be the distribution function of the time since the last arrival when observing the process at a departure point. The key to seeing this is to note that the last arrival does not necessarily correspond to the observed departure. 

Now with \eqref{eq:mgHG} in hand, there is an immediate scheme for conducting non-parametric inference of $G(\cdot)$ (and $\lambda$) under the (IO) observation scheme. By observing the sequence of arrivals and departures, we can construct the empirical distribution function that estimates $H(\cdot)$, and also obtain an estimate for $\lambda$ based on the arrival rate. Then \eqref{eq:mgHG} can be used to find an estimate of~$G(\cdot)$.

This general idea was revived and extended by Blanghaps et. al. in a more recent paper, \cite{blanghaps2013sojourn},  where the distribution of the $r$th latest arrival,  $H^{(r)}(\cdot)$,  is considered. The relation between $G(x)$ and  $H^{(r)}(\cdot)$ is given by:
$$
H^{(r)}(x)=1-\big(1-G(x) \big)e^{-\lambda x}\, \frac{(\lambda x)^{r-1}}{(r-1)!}-\sum_{j=0}^{r-2}\frac{e^{-\lambda x}(\lambda x)^j}{j!}.
$$
As shown in \cite{blanghaps2013sojourn}, the improvement in estimating $G(x)$ through  $H^{(r)}(\cdot)$ is considerable when $\rho$ is greater than 1.

The paper \cite{grubel2011matchmaking} by Gr\"{u}bel and  Wegener, treated the same problem, but there the authors seemed to not be aware of the \cite{brown1970m} result (and idea). Hence in Section 2 of that paper, they analysed the problem using the concept of {\em matchmaking} (guessing what departure maps to what arrival). They developed a method for matchmaking for the case in which the distribution of the sojourn times is either exponential, log-convex or  log-concave. For the last two cases they showed that this match is unique. That paper also goes a bit further to provide an hypothesis test for determining if the service times in M/G/$\infty$ are exponential. 
 
\paragraph {Infinite Server Queues Under (DI):} In contrast to many other queueing systems, the M/G/$\infty$ queue has explicit expressions for the joint distribution of $Q(t_1),\ldots,Q(t_n)$. This allows one to carry out effective inference in the discrete intervals (DI) observation scheme. In \cite{goldenschluger2016nonparametric} 
Goldenshluger constructed a nonparametric estimator based on discrete observations which exploits a relationship between the derivative of the covariance function and the distribution $G$. The study of nonparametric inference for this queue was then extended in \cite{goldenshluger2018m} where Goldenshluger considered the variant where the  arrival and departure epochs are registered without knowledge of the epoch type. That paper contains further results and comparisions to previous estimators.

Extended models include the time inhomogeneous case studied in \cite{goldenshluger2019nonparametric} by Goldenshluger and Koops. Further, in a discrete time setting, in \cite{edelmann2014nonparametric}, Edelman and Wichelhaus, considered parameter estimation for two-node networks of infinite server queues with geometric arrivals and general service times.  A related paper, \cite{schweer2015nonparametric}, studied parameter estimation for discrete time G/G/$\infty$ queues. The work was extended in \cite{schweer2020nonparametric} in the context of queueing networks.

\paragraph {Observing Busy Periods:} Consider a situation where the sequence of busy periods $\{B_n\}$, as well as idle periods, is observed. This was considered in \cite{hall2004nonparametric} where a sequence of busy period observations is used to construct empirical approximations to the distribution function of the service time. A related paper is \cite{bingham1999non}.

Inference procedures for M/G/$\infty$ queues based on busy and idle periods may seem attractive from a statistical perspective, but from a practical (and/or queueing) perspective they are less useful. As an example consider an M/G/$\infty$ queue with $\lambda = 50$ (customers per minute) and $\mu = 1$ (mean service time of a customer is $1$ minute). This could model, for example, the underground crossing of a  major street mentioned above where the walking time is about $1$ minute and there are about $50$ people entering the crossing every minute. In such a case, the stationary distribution of the queue length is known to be Poisson distributed with parameter $\rho = 50$ and the stationary probability of being empty is thus $e^{-50}$. Now the time between idle periods i.e.\, $\tau = \inf\{t>0 ~|~ Q(t)=0\}$, satisfies,
\[
\E[\tau ~|~ Q(0^-) = 0, Q(0)=1] = \lambda e^{\rho} = e^{50}.
\]
Hence it is not reasonable to expect to actually collect any data in such a scenario. Thus papers that base their statistical analysis on this observation scheme essentially deal with a situation that is unlikely to occur in practice. 

\paragraph{More Related Work:} In \cite{pickands1997estimation} Pickands and Stine considered a discrete time infinite server system with geometric service times where the queue size is the only observation. They proposed an estimator for the arrival rate and the holding time distribution. Their key contribution was that they model the situation with a HMM where the hidden component was the order of arrivals and departures. They used HMM algorithms and the correlation structure 
of the process for constructing MLEs. A related line is by Brillinger \cite{brillinger1974cross} where he developed a spectral approach for estimator construction of G/G/$\infty$ models and generalizations. 

\subsection{Inference with Discrete Sampling}
\label{discrete}

In the previous section we overviewed work dealing with the (DI) observation scheme for infinite server queues. We now discuss estimation under this observation scheme for other models. In computerized applications the data often includes a full log of queueing information. However, in physical systems, periodic logging is often more sensible. 
We illustrated such an example in Section~\ref{sec:methods}, where for instance, the sample of the queue length over discrete intervals is given in \eqref{eq:sampleQ1}. 

The difficulty with discrete sampling is that unless we are considering non-interacting systems as in the previous section, the joint distribution of the samples is typically intractable. Hence research in this area often builds on approximations or modifications of the sampling scheme. As illustrated in Section~\ref{sec:methods}, if the interval between samples is very large, then we can use a crude approximation where the samples are assumed to be independent. However, one needs to keep in mind that the assumed stationarity of the system is questionable when considering large sampling intervals. 

For continuous time Markov chain models with small finite state spaces, sampled at discrete intervals, one can construct maximum likelihood estimation procedures. For instance, see \cite{bladt2005statistical}, where Bladt and S{\o}rensen established the existence and uniqueness of the MLE, and compared the use of the EM-algorithm and alternative Markov chain Monte Carlo (MCMC) based procedures. This method can be applied to small finite state Markovian models of queueing systems. However the method quickly becomes intractable as the state space grows.  

In terms of approximations, in \cite{ross2007estimation}, Ross et. al.\ considered M/M/$c$ queueing systems under the (DI) observation scheme. They approximated the process using Ornstein-Uhlenbeck diffusion approximations which work well when the number of servers $c$ is not small. They carried out MLE estimates on the approximated model. A related preprint is  \cite{mcvinishconstructing}, where McVinish and Pollett considered the method of `estimating equations', which to the best of our knowledge has not been exploited further in the context of queueing inference. 

A modification of the (DI) observation scheme is to use Poisson probing, where samples occur at times dictated by a Poisson process independent of the other system processes. For models such as the M/G/1 queue, as well as more general L\'{e}vy-driven storage systems, Poisson probing yields tractable estimators. In \cite{ravner2019estimating}, Ravner et. al. exploited the fact that the dependence structure of the workload process, sampled according to a Poisson process has closed form. Specifically, given the value of the workload process at a specified time, the Laplace transform of the workload process at an exponential future time was explicitly derived. They exploited this structure to carry out, and analyze, semi-parametric estimation of the L\'{e}vy exponent driving such queues. Further, in \cite{mandjes2019hypothesis}, Mandjes and Ravner considered hypothesis testing for such systems. Related to these papers is \cite{duffie2004estimation} by Duffie and Glynn, where the authors introduced a generalization of the method of moments for continuous time Markov chains sampled at random time intervals. Another related paper is \cite{den2017convergence}, where den Boer and Mandjes considered a general estimation problem using Laplace transforms, also with application to the M/G/1 queue.

\subsection{Inference with Queueing Fundamentals}
\label{fundamentals}
Queueing theory supports many models, each with its own properties and theoretical results. However, there are also basic fundamental properties of queues that are universal to almost any queueing model. These include Little's law, see \eqref{eq:little}, as well as general properties such as the fact that queue lengths in critically stable queues are often of the order $O\Big((1-\rho)^{-1}\Big)$, and the fact that tail asymptotics of waiting time and sojourn time distributions often have a known asymptotic form. 

Several key papers have exploited such properties for the purpose of inference and estimation. In terms of tail asymptotics, in \cite{glynn1997parametric}, Glynn and Torres considered how long the arrival process needs to be observed in order to accurately estimate the long-run fraction of time that the workload exceeds a given level. Their conclusion appears to hold regardless of whether the arrival process exhibits complex dependencies or not. In \cite{glynn28estimating}, Glynn and Zeevi, established logarithmic consistency and studied the efficiency of tail based estimators. In \cite{duffy2009estimating}, Duffy and Meyn, considered Lindley recursions similar to \eqref{eq:lindleyR}, and studied their estimation properties via a large deviations analysis.

In terms of Little's law, there has also been significant work. In many situations one may  observe either the queue length trajectory, or the sojourn times of customers, or both. Little's law ties the expected value of these two quantities and hence whenever we can use one quantity for estimation, we can also use the other. As an example, refer back to the M/D/1 queue and the waiting time based estimator, \eqref{eq:md1Est2}. In that case, the sample mean of the waiting time, $\overline{W}$ was observed. However, consider a situation where instead we observe the time-average of the number of waiting customers,
\[
\overline{Q}_r = \frac{1}{T} \int_0^T (Q(u)-1) {\mathbbm 1} \{ Q(u) \ge 1\}  \, du.
\]
Now by Little's formula we expect,
\[
\overline{Q}_r \approx \lambda \overline{W}.
\]
This motivates writing down an estimator for $\rho$ via an adaptation of \eqref{eq:md1Est2},
\[
\frac{\overline{Q}_r}{\rho/m} = m \frac{\rho}{2(1-\rho)}.
\]
Now solving the quadratic equation for $\rho$ and choosing the non-negative solution, we obtain the estimator,
\begin{equation}
\label{eq:estQr}
\hat{\rho} =  \sqrt{\overline{Q}_r(2+\overline{Q}_r)} -\overline{Q}_r.
\end{equation}

A comparison of the estimator \eqref{eq:estQr} with the estimator \eqref{eq:md1Est3} or the similar estimator \eqref{eq:lambdaMD1est}, indicates that there are multiple methods to estimate the same quantity. With infinitely many samples, these methods are equivalent, however in general there is room to investigate the statistical properties of such competing estimation schemes. 

Results in this spirit were analyzed in depth in \cite{glynn1989indirect} by Glynn and Whitt. In that paper, the authors extended variance reduction results by Carson and Law, \cite{carson1980conservation}, and investigated trade-offs relating to Little's law based estimation. They focused on estimation of the means of the endogenous processes, $Q_r$ and $W$ (or respectively $Q$ and the sojourn time process). They considered the arrival rate $\lambda$ as either known or unknown and they further dichotomised between what they call the ``direct estimation'' and the ``indirect estimation'' case. In the former, the mean of $Q$ is estimated directly by a time average of the observed queue length. In the latter Little's law makes use of the sample mean for $W$ together with $\lambda$ or an estimator for it, to estimate the mean of $Q_r$ (or $Q$). The results of this paper relied on the authors' earlier work in \cite{glynn1986central}, as well as other joint papers, which established a joint functional central limit theorem based on Little's law, which describes the weak convergence of both the queue length estimator and the waiting time estimator, to an appropriate limiting diffusion process. The results in \cite{glynn1989indirect} indicate that an indirect estimator is more efficient than a direct estimator in cases where the interarrival and waiting times are negatively correlated. 

This line of research has been further extended in \cite{nozari1988estimating} in the context of manufacturing performance analysis. Further in \cite{glynn1993estimating},  Glynn et. al.\ established a martingale central limit theorem which they then used to construct confidence intervals for estimators and perform statistical tests. In more recent work, in \cite{kim2012estimating}, Kim and Whitt surveyed previous results and refined the Little's law based estimators to handle certain cases such as removing bias due to interval edge
effects. See also \cite{kim2013statistical} for a treatment of the time varying version of Little's law.

A different line of research that we would like to highlight here is the detection of stability or instability of queueing systems. To date not much work has been done towards this direction but one notable publication is 
\cite{mandjes2017detecting} where the authors deal with Monte Carlo simulation of systems for detecting stability or instability.

\subsection{Queue Inference Engine Problems}
\label{QIE}

A paper published in 1990, \textit{The queue inference engine: deducing queue statistics from   transactional data}, by Larson  \cite{larson1990queue}, opened up a research direction associated with retrospective state estimation of a queue based on {\em transactional data}. This is what we call the (T) observation scheme, see Table~\ref{tab:obs}. The key idea is that only observations of ordered service entry and service completion times are available. Under this observation scheme the cumulative number of departures for $t \in [0,T]$, $D(t)$ is observable, however, the cumulative number of arrivals $A(t)$ is not. This occurs in applications such as automatic teller machines where the queue of customers waiting for the machine is not observable but the transaction record is logged.

The only statistical assumption is a (homogeneous) Poisson arrival process and no further assumptions on the service processes. That is, the inference works for models such as M/G/$c$ as well as more complicated models with Poisson arrivals. As was shown in \cite{larson1990queue}, smart recursions utilizing the uniform order statistics property of Poisson processes, can be utilized to infer retrospective mean queue length trajectories and other quantities, during congestion periods (duration during which all servers are busy), only based on transactional observations during that period.

\begin{figure}
\centering
\includegraphics[scale=0.7]{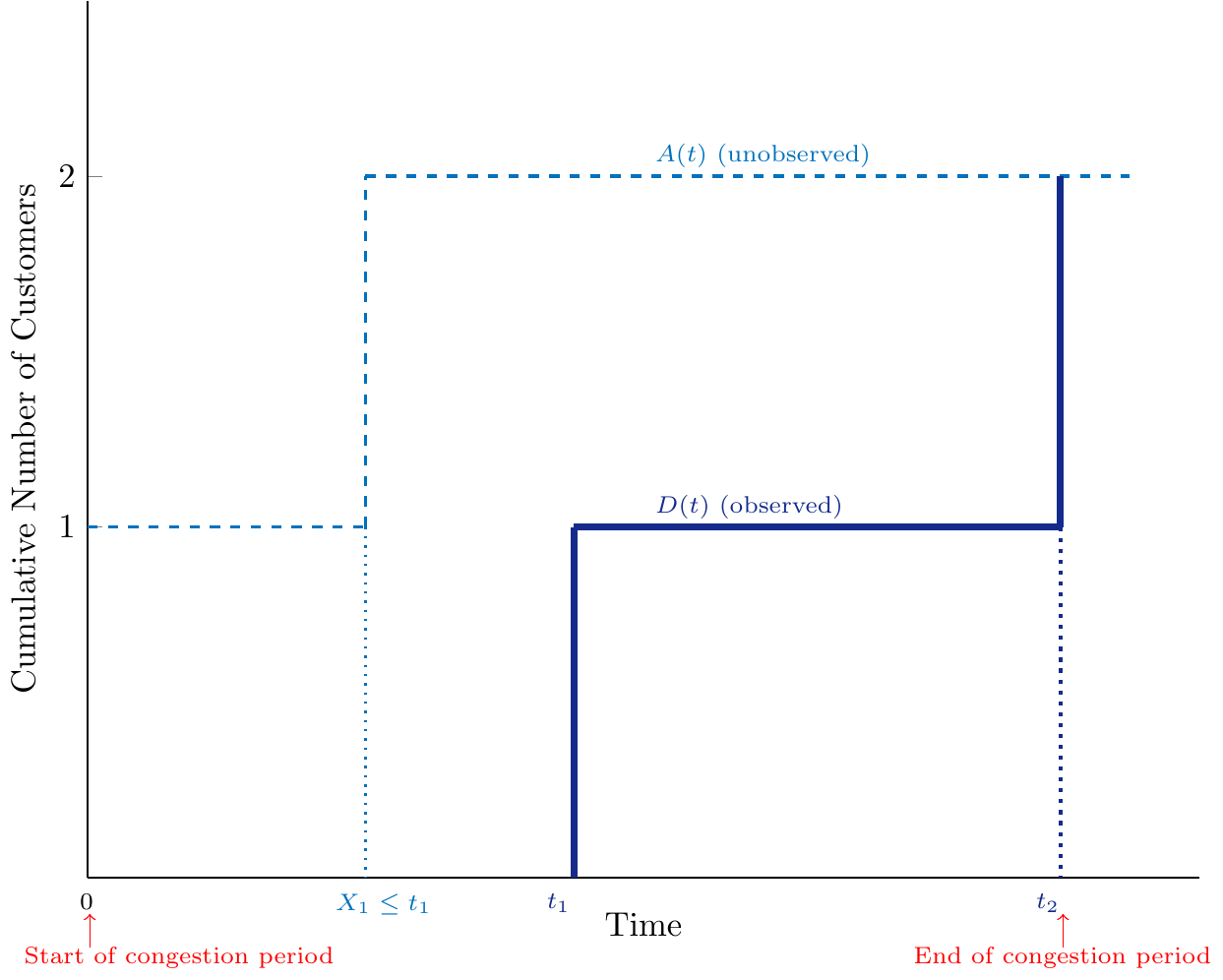}
\caption{\label{fig:QIE2}
{\small The cumulative number of customers during a congestion time. Here, $A( t)$ is the cumulative number of arrivals and $D(t)$ is the cumulative number of departures during the congestion period. The arrival time of the $i$-th customer to enter the queue and the departure time  of the $i$-th customer served  are denoted by $X_i$ and $t_i$, respectively.
}}
\end{figure}

As an elementary illustrative example see Figure~\ref{fig:QIE2} which assumes a single server case. Here the congestion period begins at time $0$ and then a completion of a transaction is recorded (a customer departing) at time $t_1$. Further, the server becomes idle at time $t_2$ (and this information is recorded in the transactional data).  Based on this information, it is clear that there were $2$ arrivals for this specific congestion period with the first arrival at time $0$ and the second at some unknown time $X_1 \in (0,t_1)$. 

Under the Poisson arrival assumption, we know that $X_1 \sim \mbox{uniform}(0,t_1)$. This then allows us to deduce the expected number of customers in the waiting room, as
\begin{equation}
\label{eq:simpleQIE}
\hat{Q}_r(t) = \E[
Q_r(t)] = \begin{cases}
\frac{t}{t_1} & t \in [0,t_1), \\
0 & t \in [t_1,t_2].
\end{cases}
\end{equation}
The computation of the estimator $\hat{Q}_r(\cdot)$ is based on the transactional observations and can only be made at time $t_2$ once it is known that the congestion period has ended. 

\begin{figure}
\centering
\includegraphics[scale=0.7]{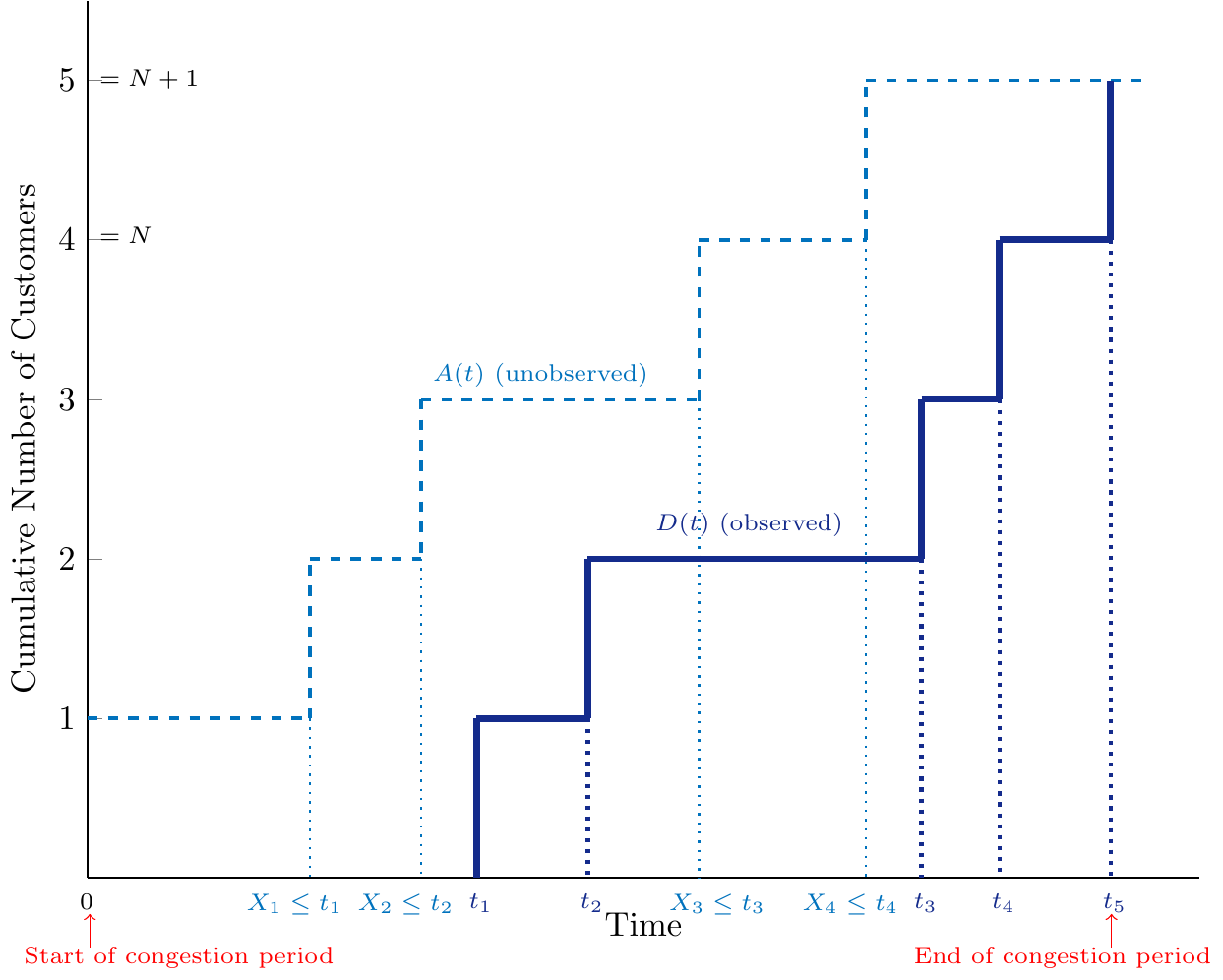}
\caption{\label{fig:QIE1}
{\small Trajectories of $D(t)$ (observed) and $A(t)$ (unobserved) during a congestion period with $N=5$ customers.}}
\end{figure}

In a more complex situation, there will be multiple transactions recorded during a congestion period. We illustrate such a situation in Figure~\ref{fig:QIE1}. At time $t=0$ an arrival occurs and the server's state changes from idle to busy; hence the congestion period starts. 
From transactional data, the system exhibits both service completion and service commencements at times $t_1$, $t_2$, $t_3$, and $t_4$. Further, there are customers waiting to be served at least at times $t_1^-$, $t_2^-$, $t_3^-$, and $t_4^-$. During the congestion period, a total of $N=4$ customers are delayed in the queue and at time $t_5$, the transactional data indicates a service completion but no service commencement, thus ending the congestion period and allowing the server to idle. It is evident that the service completion times within a congestion period impose a set of inequalities on the arrival times of other customers who waited in queue:
\begin{equation}
\label{eq:arrivalConstraints}
X_1 \le t_1,
\ldots
\quad X_N \leq t_N.
\end{equation}

The queue inference engine allows us to compute an estimator, $\hat{Q}_r(\cdot)$ based on this data with the main idea being a recursive computation that uses the uniform arrival property for Poisson arrival processes, similarly to the simple estimator in \eqref{eq:simpleQIE}, as well as the inequalities in \eqref{eq:arrivalConstraints}.  Details are in \cite{larson1990queue}.

Larson's \cite{larson1990queue} paper developed a system linear equations that can be solved at the end of each congestion period for computing $\hat{Q}_r(\cdot)$. Note that given $n$ arrivals during a congestion period, the computation time needed to obtain mean queue lengths is of order $O(n^5)$. See \cite{larson1990queue} for numerical examples, compared to simulation.

Following Larson's paper, a variety of research papers generalized the basic idea and presented improvements to the computation time. In \cite{bertsimas1992deducing}, Bertsimas  and Servi improved the computational complexity and generalised the results to cases of non-homogeneous and renewal arrival processes. Note that for the homogeneous Poisson arrival process, the unordered arrival times are independent and uniformly distributed, but for the non-homogeneous case with the arrival rate $\lambda(t)$,  the unordered arrival times are i.i.d. with a probability density proportional to $\lambda(t)$. In \cite{daley1992exploiting} Daley and Servi, extended the result to the case of Erlang-$k$ interarrival times. They also considered the cases of having finite buffers and the real-time estimation problem where the arrival rate is known. Further, in \cite{daley1993two}, they considered systems where arriving customers can balk when the queue length is beyond a given threshold and the balking probability is constant. The results of that paper were extended by Jones in \cite{jones1999inferring} where the  balking probability increases to $1$ when the queue length approaches~$\infty$.

Meanwhile, Jones and Larson \cite{jones1994efficient} suggested additional algorithms with $O(n^3)$ computational complexity, improving the algorithms of \cite{larson1990queue}.
In \cite{daley1997estimating}, Daley and Servi presented an additional algorithm by omitting queue lengths with low probabilities. Later on, in \cite{daley1998moment}, the result was extended to customer balking and reneging. In \cite{dimitrijevic1996inferring} Dimitrijevic developed further algorithms which under special cases have complexity as low as $O(n^2)$. Moving onto queueing networks, in \cite{mandelbaum1998estimating}, Mandelbaum and Zeltyn applied the queue inference engine idea. This is one of the few papers in this survey that deal with complex queueing networks (as opposed to single pass queueing systems). 

In all of these papers, calculating  the likelihood of a congestion period is the most difficult task. This difficulty is due to the fact that the likelihood should be integrated over the all realisations of the unobserved arrival process and the number of terms in this sum increases exponentially with the number of departures. In \cite{fearnhead2004filtering}, Fearnhad considered applying a likelihood recursion to test the likelihood efficiency of the estimator when used in the M/G/1 and E$_k$/G/1 cases. Here, the only observed variables were the inter-departure times. 

In \cite{frey2010queue}, Frey and Kaplan considered the case of periodic reporting data, where the arrivals follow a Poisson process with period-specific arrival rates and the data is the number of departures during each period. However, the results of this paper were challenged by Jones in \cite{jones2012remarks} where he showed that queue inference cannot be carried out without knowing service start or stop times. Further, in \cite{jones2012remarks} Jones,  presented an extension of the analysis.

In \cite{park2011analysis} Park et. al. presented a new complementary variant of the QIE problem. They considered the case where the number of servers is unknown and exact inferences about queueing and service times come from the arrival and departure times. Then, in \cite{heckmuller2011reconstructing},  Heckm{\"u}ller and Wolfinger, considered the case of the G/D/1 queue for inference about characteristics of the arrival process from transactional data. 
\subsection{Bayesian Approaches}
\label{Baysian}

A Bayesian approach to inference treats unknown parameters as random variables and the inference procedure is a process of refining the distribution of these parameters. We begin with a prior distribution on the parameter values and once data is collected, Bayes' formula yields a posterior distribution. This posterior distribution and functionals of it are the main outcome of the inference.

In setting up a Bayesian estimation problem, prior distributions are often parameterised themselves by hyper-parameters, and in certain cases, the resulting posterior distribution also has a parametric form. Such cases arise when the prior is a conjugate for the likelihood model, that is, when the distribution of the posterior has the same parametric form as the distribution of the prior. When there is not a nice parametric form for the posterior, computational methods are required, especially when the posterior distribution is high-dimensional with an intractable normalization constant. Common methods include Monte Carlo Markov chains (MCMC), as well as many modifications and adaptations such as approximate Bayesian computation (ABC) which is used in case of an intractable likelihood. See for example, \cite{bolstad2016introduction} for more details about Bayesian inference at large.

The Bayesian approach usually applies to queueing systems where in addition to inference, prediction is of interest. Example applications include internet traffic analysis and risk theory. Let us return to a very elementary example of the M/D/1 queue and as with previous examples assume that the service duration $m$ is known and  the unknown arrival rate $\lambda$  is the parameter of interest. Consider now the (F) observation scheme, see Table~\ref{tab:obs}, where a full queue trajectory is observed, however it is only the arrival process which is of interest. Now the data collection is over $n$ periods, each of duration $T$, and the data is a sequence $\{x_1,\ldots,x_n\}$ where $x_i$ is the number of arrivals in period $i$. In this case, the likelihood model is that the distribution of the number of arrivals during a period is Poisson with parameter $\lambda T$. 

A common choice that works well with the Poisson likelihood is a Gamma prior with hyper-parameters $\alpha$ and $\beta$ as shape and rate parameters, respectively. In this case, it is an elementary Bayesian calculation to show that gamma distribution is a conjugate prior.

To see this note that the posterior is proportional to the product of the likelihood and the prior and hence,

\begin{equation}
\label{eq:gammaPoissonBaysUpdate}
\begin{array}{rl}
f(\lambda ~|~ \text{data} ) ~ \propto ~& \Big(\prod_{i=1}^n e^{-\lambda T} \frac{(\lambda T)^{x_i}}{x_i!} \Big) 
\frac{\beta^\alpha}{\Gamma(\alpha)} \lambda^{\alpha-1} e^{- \beta \lambda} \\ \\
~ \propto ~& e^{-n\lambda T}~~ (\lambda T)^{\sum_{i=1}^n x_i} ~~ \lambda^{\alpha-1}~~ e^{- \beta \lambda } \\ \\
~ \propto ~& e^{-n\lambda T}~~ \lambda^{\sum_{i=1}^n x_i} ~~ \lambda^{\alpha-1}~~ e^{- \beta \lambda } \\ \\
~ = ~ & \lambda^{\alpha + (\sum_{i=1}^n x_i) - 1} \,e^{-\lambda(\beta+nT)}, \\
\end{array}
\end{equation}
which is proportional to the density function of a gamma distribution with shape parameter $\alpha + \sum x_i$ and with rate parameter $\beta + n T$.

This example is of course just performing inference for the rate of the Poisson arrival process although it is posed here in the context of an M/D/1 queue. A similar line of reasoning has been employed in a significant body of literature dealing with Bayesian inference for queues. For example see early work of Armero et. al. in \cite{armero1985bayesian}, \cite{armero1994bayesianInf}, \cite{armero1994bayesian}, and \cite{armero1994prior} for obtaining the posterior distribution of the traffic intensity, waiting time, number of customers, and length of idle and busy period for an M/M/1 queueing system. Later similar work, focusing on specific phase type  service times, can be found in \cite{ausin2004bayesian} and \cite{insua1998bayesian}. See also  \cite{ramirez2010bayesian}, where additional specific distributions  amenable for efficient Bayesian inference are employed and \cite{mcgrath1987subjective} and \cite{mcgrath1987subjectiveII} where considerations of the subjective Bayesian paradigm for queueing inference are discussed.

An alternative computational line of research includes \cite{sutton2011bayesian} by Sutton and Jordan where Bayesian inference for general queueing networks and service mechanisms was studied. Here the queue is generally viewed as a transformation mechanism between exogenous processes and endogenous processes (although the authors don't use this terminology). They considered a variety of mechanisms and policies and presented an overview of the application of Bayesian inference for queueing networks where simulation of the queueing processes is part of the posterior procedures to sample latent variables. The computational procedures make use of the slice sampler, \cite{neal2003slice}.
The computational paradigms introduced in \cite{sutton2011bayesian} have influenced several other works in the computer science and Bayesian statistics communities, such as \cite{wang2016maximum} where closed queueing networks were considered. 

The Bayesian paradigm also extends to empirical Bayesian approaches as discussed for queues in \cite{thiruvaiyaru1992empirical} and to more recent work dealing with non-parametric Bayesian approaches. 
A notable paper in this direction is \cite{conti1999large} focusing on discrete time queues where the inference is for the service time distribution.
\subsection{Online Prediction}
\label{predict}

In this paradigm, we observe some of the endogenous processes up to a  given time and make prediction about future values. A common application is {\em delay prediction} where the queue length process or workload process is observed and used for predicting the waiting time of arriving customers. Using delay predictions to make {\em delay announcements} is common in call-centers and other service operations. In certain cases some of the model parameters are known.

Most of the literature considers the mean square error (MSE) criterion, under which the best predictor is the expected value. As an example, consider a GI/G/1 system with known arrival rate $\lambda$ and known mean service time $m$. Assume that at time $t_0$ we observe $Q(t_0) = q_0>0$. We may then require predictors for the waiting time of a customer arriving at time $t_0$ (not included in the count $q_0$) or for functions of the future queue length, $f\big(Q(t_0+u)\big)$. Such predictors also include the waiting time of future customers that  arrive at $t_0 + u$.

Under a FCFS policy, predicting the waiting time of
a customer arriving to find $q_0$ customers already in the system 
is straightforward. The expected service time of each of those
yet to commence service  is $m$ and hence the expected delay is,
\[
(q_0-1)m + R,
\]
where $R$ is the residual service time of the customer currently in service. The value of $R$ may either be observed, or estimated. In a case such as the GI/M/1 queue, the expected delay is simply $q_0 m$ due to the memory-less property of the exponential distribution. Further, it is straightforward to provide quantiles or other measures of the delay time, as the waiting time of the customer is Erlang (Gamma) distributed. 

If we are looking for predictors for $f\big(Q(t_0+u)\big)$, then explicit expressions require more stringent assumptions. For example, in an M/M/1 queue, $Q(t_0)$ describes the full state information and 
the predictor that minimizes the MSE is,
\[
\widehat{f}(Q(t_0 + u)) = \E[f\big(Q(u)\big) ~|~ Q(0) = q_0] = \sum_{j=0}^\infty \, f(j)\, p_{q_0j}(u),
\]
where $p_{ij}(u)$ is the transition probability of a birth-death continuous time Markov chain, from state $i$ to state $j$ in $u$ time steps. In the case of M/M/1, expressions involving Bessel functions for $p_{ij}(\cdot)$ are known, \cite{cohen1982single}, and hence in principle closed form predictors can be computed. However in more general models, predictors for $f\big(Q(t_0+u)\big)$ quickly become intractable and hence approximations are involved.

The classic literature dealing with such cases includes \cite{stanford1983optimal} where transition probabilities for the GI/M/1 embedded Markov chain are used, \cite{woodside1984optimal} where extensions to multi-server GI/M/$c$ queues are considered, and \cite{pagurek1988optimal} where predictors associated with the M/G/1 queue are considered. In these papers, explicit transition probabilities of certain endogenous processes were used along with the embedded  Markov chain structure of GI/M/$c$ and M/G/1 type queues. In general, there does not appear to be a mechanism for generalizing this type of analysis beyond GI/M/$c$ and M/G/1. That is systems such as M/G/$c$ or GI/G/1 queues or more complex systems require a different set of tools.

For more general settings, one can consider approximations. A general entry point focused on operations management of call centres is  \cite{whitt1999predicting} deriving  predictors for the waiting time of customers currently in the system. The analysis focuses on multi-server systems with multiple customer classes. Further work was carried out in \cite{ibrahim2011wait} where the realistic scenario of time-varying demand and time-varying service capacity was considered. In such cases, fluid approximations were employed to derive several types of predictors. See also~ \cite{thiongane2016new}. 

\subsection{Implicit Models}
\label{Info-mining}

In many of the paradigms described above queueing models are explicitly used to model real life situations and data is used for parameter estimation or state prediction. However, queueing models may also be used implicitly, without requiring an ``exact fit'' between model and reality. Towards that end, several different research directions have been pursued. One direction is the application of {\em information mining} based on {\em event logs} for creating queueing models directly from the data. Another direction is {\em grey-box modelling} where queueing like processes are used to describe the data, without requiring an exact fit. 

\paragraph{ Information mining:} The general idea here is to use an extensive event log dataset to dynamically create queueing models that describe the underlying processes. This is quite different from classical modelling where the modeller observes the process and suggests a mathematical model. This idea has been explored in a series of recent papers. In \cite{senderovich2014queue} Sendrovich, Weidlich, Gal, and Mandelbaum use the developed field of {\em business process mining} based on event logs, see \cite{van2004workflow}, for queues. They adapted ideas from this field to queues and developed the method of {\em queue mining}. In \cite{senderovich2015discovering} the work was extended to handle the queue mining paradigm in view of partial information. In \cite{senderovich2015queue} customers with different priorities were incorporated as part of the queue mining process. Further, in \cite{senderovich2015discovery} a resource-driven perspective was employed with an application to an outpatient clinic.

\paragraph{ Grey-box models:} There are several classes of stochastic processes that are often used in explicit queueing models. These include birth and death processes and other structured Markov chains. One may develop {\em statistical queueing models} which are based on similar underlying processes but 
do not attempt to utilise a mechanistic relationship between models for the exogenous and endogenous processes. 
As an analogy consider time-series models where  common stochastic processes, such as 
autoregressive integrated moving average (ARIMA) models, are used without an explicit description of how the underlying random variables are related to the physical world. This idea can be used with birth and death processes or with any other queueing based stochastic process in the hope that the queueing-like stochastic process can model queueing phenomena well.

In \cite{dong2015stochastic} Dong and Whitt considered a stationary birth and death process fitted to a sample path of an arbitrary queueing system. General birth and death parameters were allowed. This differs from an explicit queueing model such as M/M/$c$ where the birth rate is assumed to be constant $\lambda$ and the death rate at level $k$ is $\mu k$ for $k \le c$ and $\mu c$ for $k > c$. In the latter scenario an exact queueing model could be fitted to estimate the parameters (parameters are $\lambda$, $\mu$, and $c$) whereas in the Grey-box approach of Dong and Whitt, an arbitrary birth and death process allows us to compensate for potential model misspecification. A similar approach was applied to health-care data in \cite{au2009predicting}. Another grey-box type paper is \cite{zhang2002workload} where queueing networks are approximately fitted to network data.

The fitting of birth and death processes is also interesting in its own right and as shown in \cite{whitt2012fitting}, different fitting methods are possible. See also \cite{dong2015using} dealing with time-varying periodic queues.

\subsection{Control, Design, and Uncertainty Quantification}
\label{control}

Most of the paradigms surveyed above deal with parameter or state estimation. However related problems deal with how to control queueing systems in the presence of uncertainty, how to design such systems, and how to deal with uncertainty quantification when carrying out such control or design. There is an extensive literature for control, design, and architecture selection for queueing systems. However, the literature mainly focuses on cases which assume that the probability laws of arrival and service processes are precisely known and the state of the system is fully observable.

In the realm of stochastic control, there are two general paradigms for dealing with such uncertainty. In one paradigm, a controller wishes to optimally control a system in which parameters are not known. This is sometimes called adaptive control. The field of reinforcement learning, suggests a variety of methods for dealing with such a setting. An alternative case is that in which the state observation is not fully available. The field of partially observable Markov decision processes (POMDPs) deals with this setting. To the best of our knowledge, in the specific context of queueing control, both of these areas have not received extensive attention.

An early paper dealing with adaptive control of queues was \cite{hernandez1983adaptive} where for an M/G/1 queueing system with an unknown arrival rate and an average cost criterion, the controller chooses the service rate  to minimize long term costs. In \cite{krishnasamy2018learning}, the celebrated $c \mu_i$ scheduling rule was analyzed in the case where the service rates $\mu_i$ are estimated online. In this case a regret based analysis was performed.
As for POMDPs, some recent work was presented in \cite{asanjarani2019role} where the interaction between partially observable queues and stability was explored. Also related is \cite{armony2005impact} where supply systems were considered and the effect of not being aware of duplicate orders is analyzed. Beyond these adaptive control and POMDP papers, we are not aware of further significant work.

In addition to control, there is the problem of how to design queueing systems. This often refers to off-line specification of quantities such as the number of servers, server rate allocation and the queueing discipline. In contrast, control of a system typically considers the problem of  on-line decision making based on state measurements or estimates. An interesting aspect dealing with design arises when parameter uncertainty is present. As an illustration of the trade-offs inherently involved, in \cite{dinh2014architecture} the authors considered a single-pass loss-less queueing system in steady-state with an unknown arrival rate. They analyzed several trade-offs dealing with architecture selection for such systems.

In general, design of queueing systems is often based on performance analysis which includes computing functionals of the endogenous processes. In recent years, there has been much work on the robust evaluation of such performance measures. In this setting, parameters are assumed to not be known exactly, but rather to lie within specified uncertainty sets. See for example \cite{bandi2015robust} and references within. 

\section{Conclusion}
\label{conclusion}

The queuing theory literature spans multiple journals, dozens of books, and thousands of publications. However within that, the literature dealing with parameter and state estimation is much more limited. 
We have done our best to list this comprehensively in the annotated bibliography \cite{asanjarani2017parameter}. Our purpose in this survey is to present an up to date account of this more narrow aspect of queueing research. While our discussion is not  exhaustive, we have attempted to present a comprehensive view of the estimation paradigms that have been investigated to date. 

When attempting to classify a body of work  one approach is to consider the several dimensions that specify the problems at hand. As outlined in Section~\ref{sec:frameBackground} for parameter and state estimation in queues these dimensions include the inference activity, the models, the observation scheme, and the statistical methods and principles. We have attempted to describe the field using this viewpoint and the ten major estimation paradigms that appear in Table~\ref{Tab1} and are surveyed in the subsections of Section~\ref{sec:settingAndActivities}.

In considering the estimation paradigms outlined in Section~\ref{sec:Problems Solved} we believe that there is room for extensive further research that will connect some of the paradigms. Specifically, the joint application of inverse problem methods, Bayesian approaches, implicit models, and control of systems may be of interest.
The past decade has witnessed an explosive growth in data driven statistical learning applications. Some of the application areas that have benefited from this growth include systems of congestion, resource scarcity, and queues. It remains a challenge to connect queueing estimation paradigms with  modern machine learning applications and methods.

\vspace{20pt}

\noindent
{\bf Acknowledgment} Azam Asanjarani's and Peter Taylor's research is supported by the Australian Research Council (ARC) Centre of Excellence for the Mathematical and Statistical Frontiers (ACEMS).
Yoni Nazarathy is supported by ARC grant DP180101602. 
We are grateful to Liron Ravner for feedback on an early version of the manuscript. We also thank Ross McVinish and an anonymous referee for useful feedback. We thank Phil Pollett for contributions to an early version of the associated annotated bibliography \cite{asanjarani2017parameter}. 

\newpage
\bibliographystyle{acm}
\bibliography{papersDB, BooksChaptersDB,biblio,ThesesisDB}

\end{document}